\documentclass[aip,jcp,graphicx,reprint]{revtex4-1}
\pdfoutput=1
\usepackage{graphicx}
\usepackage{amssymb}
\usepackage{amsmath}
\usepackage{blkarray}
\usepackage{multirow}
\usepackage{natbib}
\usepackage{epstopdf}

\draft 
\begin{document}


\title{Molecular Dynamics Simulations of Solutions at Constant Chemical Potential}



\author{C. Perego }
\email[]{claudio.perego@phys.chem.ethz.ch}
\affiliation{  Department of Chemistry and Applied Biosciences, ETH Zurich,  Zurich (Switzerland)} 
\affiliation{  Institute of Computational Science, Universit\`a della Svizzera Italiana, Lugano (Switzerland)}

\author{M. Salvalaglio }%
\affiliation{  Institute of Computational Science, Universit\`a della Svizzera Italiana, Lugano (Switzerland)}
\affiliation{  Institute of Process Engineering, ETH Zurich, Zurich (Switzerland)} 
\author{M. Parrinello}
%
\affiliation{  Department of Chemistry and Applied Biosciences, ETH Zurich, Zurich (Switzerland)} 
\affiliation{  Institute of Computational Science, Universit\`a della Svizzera Italiana, Lugano (Switzerland)}



\date{\today}

\begin{abstract}
Molecular Dynamics studies of chemical processes in solution are of great value in a wide spectrum of applications, which range from nano-technology to pharmaceutical chemistry. 
However, these calculations are affected by severe finite-size effects, such as the solution being depleted as the chemical process proceeds, which influence the outcome of the simulations. To overcome these limitations, one must allow the system to exchange molecules with a macroscopic reservoir, thus sampling a Grand-Canonical ensemble. Despite the fact that different remedies have been proposed, this still represents a key challenge in molecular simulations. 

In the present work we propose the Constant Chemical Potential Molecular Dynamics (C$\mu$MD) method, which introduces an external force that controls the environment of the chemical process of interest. This external force, drawing molecules from a finite reservoir, maintains the chemical potential constant in the region where the process takes place.
We have applied the C$\mu$MD method to the paradigmatic case of urea crystallization in aqueous solution. As a result, we have been able to study crystal growth dynamics under constant supersaturation conditions, and to extract growth rates and free-energy barriers.
$ $\\
\begin{flushright}
\emph{\small J.~Chem.~Phys.~142,~144113 (2015); http://dx.doi.org/10.1063/1.4917200} 
\end{flushright}

\end{abstract}

\maketitle
\section{Introduction}\label{intro}

Physical-chemical processes taking place in liquid solution are ubiquitous, and are at the heart of a wide variety of applications. Electro-chemical reactions, surfactants adsorption, self-assembly, crystal nucleation and growth, are just a handful of such applications. However, in many cases, a detailed description of the molecular mechanisms at play during these processes is still lacking. 

Molecular Dynamics (MD) represents a powerful method for investigating such processes with atomistic detail. However MD suffers from many limitations such as the relatively small time and size scales that can be simulated.  
With the presently available computational resources, classical MD calculations, based on empirical potentials, can typically study systems of size up to $10^{4}\div10^{6}$ atoms. Such size limitations are particularly dramatic in the simulation of phase transformations, as in the paradigmatic case of crystal growth from solution. In such a case, while the crystallization proceeds, the solution is depleted, thus changing its chemical potential and affecting the growth process itself.  
For this reason MD results require seizable finite-size corrections before being compared with the experimental results, which involve much larger system sizes  \cite{SalvalaglioPNAS2014,SchmelzerJNCS2014,GrossierCGD2009}. 

In general, the numerical approach to prevent such finite size problems is that of sampling configurations in the Gran-Canonical (GC) ensemble, namely simulating a system in contact with an external reservoir, which maintains the chemical potential constant. However, the classical methods for GC sampling \cite{YauJCP1994,PapadopolouJCP1993,LoJCP1995,CaginMP1991,LynchJCP1997} require on-the-fly insertion and removal of particles. This constitutes a crucial obstacle for chemical processes in solution, since the probability of a successful particle insertion in a dense fluid is extremely low \cite{DelgadoBJCP2003}, preventing efficient numerical studies. 

Recently, the development of adaptive resolution simulation methods AdResS (see e.g.~Ref.~\onlinecite{WangPRX2013} and references therein) and $H$-AdResS \cite{PotestioPRL2013,PotestioPRL2013b} has lead to an alternative path to MD simulation in the GC ensemble. In these methods the region of primary interest is described with atomistic detail, while its surroundings are treated through a lower resolution model. It has been shown that the low resolution region can act as an external molecule reservoir, enforcing the sampling of the GC ensemble in the atomistically resolved region. Moreover, in this low resolution region, particle-insertion or swapping techniques can be applied more efficiently\cite{MukherjiMac2013}.

Here, inspired by this approach, we use an analogous volume subdivision and implement a method that addresses the problem of solution depletion in MD simulations. To be more definite, we focus our attention on the simulation of crystal growth from solution, which lately has received much attention \cite{PianaNat2005,PianaJACS2006,CheongCGD2010,AnwarAC2011,SalvalaglioJACS2012} and for which the pitfalls of limited size numerical modeling have been underlined \cite{SalvalaglioAC2013}. Nonetheless we stress that the present method can be applied to many other systems.

In our scheme the region containing the growing crystal and its immediate surroundings is maintained at constant solution concentration, while the remainder of the simulation box acts as a molecular reservoir. In this way we are able to study the crystal as it grows or dissolves in contact with a solution at constant chemical potential. 

The paper is organized as follows: first, in Sec.~\ref{mustat}, our method is introduced and described in detail. After that we apply it to the crystallization of urea in aqueous solution. The technical details of the calculations are presented in Sec.~\ref{simsetup}. Then, in Sec.~\ref{growth}, the simulation results are illustrated and discussed. Finally, Sec.~\ref{concl} is devoted to the conclusions and possible perspectives of our work.

\section{Constant Chemical Potential MD}\label{mustat}
\begin{figure}[] 
\centering\includegraphics[scale=0.3]{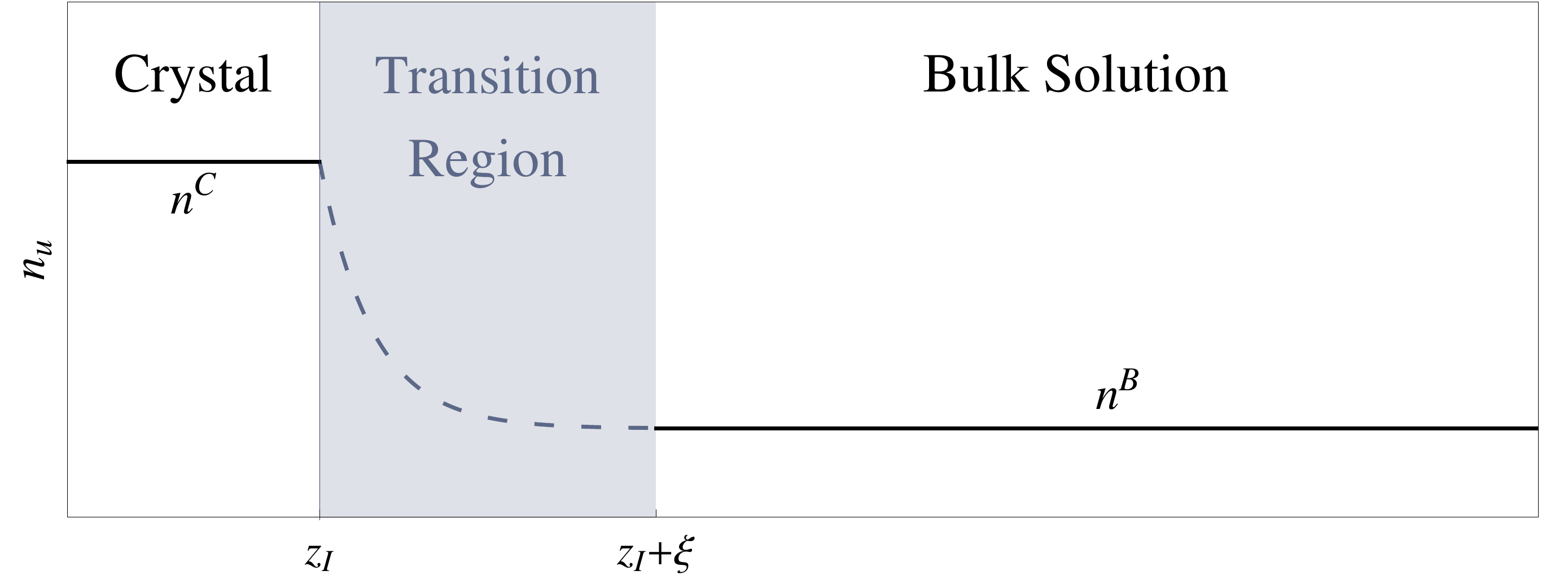}\\  
\caption{\label{scha} Qualitative behavior of the solute local density in the vicinity of a planar crystal surface. The blue, vertical line indicates the solid-liquid interface position at $z=z_{\mathrm{I}}$. The light blue shaded area corresponds to the TR, defined in the text. Within this region, the dashed line is explanatory and non-representative of the realistic $n_{\mathrm{u}}$ behavior.}\end{figure}
In the present section we describe in detail our new method, named Constant Chemical Potential MD (C$\mu$MD), which allows to perform MD of chemical processes in a solution at constant chemical potential. 
In order to illustrate the method we focus on the problem of crystal growth in a binary solution, while in principle any first order phase transition can be considered. Moreover, for the sake of simplicity, we consider a planar crystal-solution interface. Extension to other geometries is possible, although rather more complicated.

In Fig.~\ref{scha} we show a qualitative description of the local solute density $n_{\mathrm{u}}$ in the neighborhood of the crystal-solution interface, located at $z_{\mathrm{I}}$. To the left of the interface there is the crystal region, characterized by the solid density $n^{\mathrm{C}}$. On the other side of $z_{\mathrm{I}}$ we find a Transition Region (TR), of length $\xi$, in which the growth process takes place. The extension of the TR is such that for $z>z_{\mathrm{I}}+\xi$ the density reaches its bulk value $n^{\mathrm{B}}$.  
The $n_{\mathrm{u}}$ profile within the TR depends on the kinetics of crystallization and on the diffusivity of the liquid, and cannot be easily guessed a priori. Here we are implicitly assuming the considered transition does not involve macroscopic correlation lengths, so that $\xi$ is finite.

During crystal growth or dissolution, the solid-liquid interface moves together with the TR. In a macroscopic system $n^{\mathrm{B}}$ remains unchanged and the density at the boundaries of the TR is constant, leading to a stationary growth process. This is in contrast with the behavior of a finite-size system, in which $n^{\mathrm{B}}$ varies due to the limited number of molecules.
\begin{figure}[] 
\centering\includegraphics[scale=0.3]{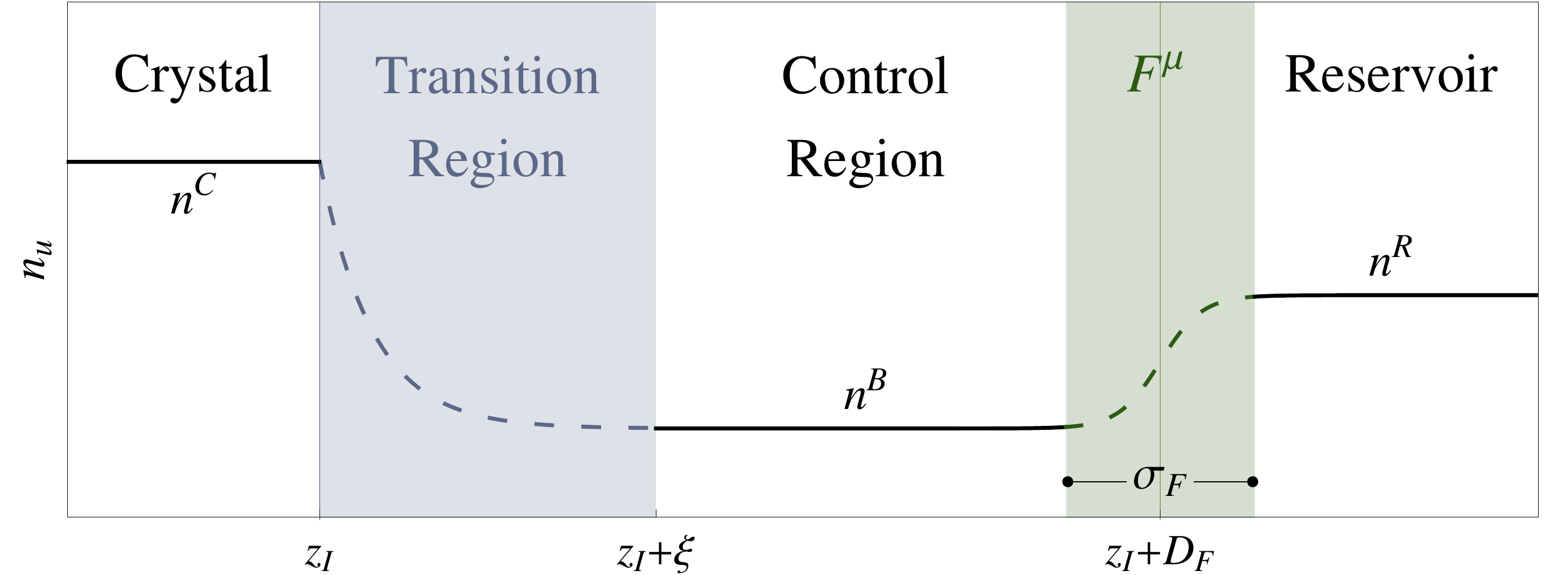}\\  
\caption{\label{schb} Same profile displayed in Fig.~\ref{scha}, with the addiction of an external force $F^{\mu}$ applied at $z=z_{\mathrm{I}}+D_{F}$ (green vertical line). The green shaded area corresponds to the transition region connecting the CR and the reservoir. Also in this case the dashed profiles are purely explanatory. }\end{figure} 

Our method aims at restoring this stationarity condition, enforcing a constant solution composition in a Control Region (CR) in contact with the TR, while the rest of the solution volume is used as a molecule reservoir, as shown in Fig.~\ref{schb}. To control the solution density, an external force $F^{\mu}$ is applied at a fixed distance $D_{F}$ from the moving crystal interface. $F^{\mu}$ acts as a membrane, regulating the flux of molecules between the CR and the reservoir, in order to maintain the former at a constant concentration. If we assume that the effect of the external force is felt in a region of size $\sigma_{F}$, the CR is located between $z_{\mathrm{I}}+\xi$ and $z_{\mathrm{I}}+D_{F}-\sigma_{F}/2$. In order not to affect the growth process the CR should be larger than the typical correlation lengths of the solution.
We stress the fact that the force is applied at a fixed distance from the solid-liquid interface, to maintain a stationary solution environment in the proximity of the growing crystal interface. 

As the crystallization proceeds the reservoir is depleted, so that the chemical potential can only be maintained for a limited amount of time $\tau_{F}$.
In general, when applying the C$\mu$MD technique to some molecular process, this time limit must be taken into account, and the simulation box has to be properly engineered so that $F^{\mu}$ is able to act efficiently during the timescale of interest. 

We write the external force $F^{\mu}$ as:
\begin{equation}
\label{Fb}
F^{\mu}_{i}(z)=k_{i}(n^{\mathrm{CR}}_{i}-n_{0i})G\left(z,Z_{F}\right),
\end{equation}
where $i$ labels the solute or solvent species, $k_{i}$ is a force constant, $n_{0i}$ is a target density and $G$ is a bell-shaped function which is nonzero in the vicinity of the force center $Z_{F}=z_{\mathrm{I}}+ D_{F}$. Eq.~(\ref{Fb}) defines a harmonic-like force, localized at a fixed distance $D_{F}$ from the solid, which acts on the solution molecules to compensate the deviations of the instantaneous CR density $n^{\mathrm{CR}}_{i}$ from $n_{0i}$.

If $N_{i}$ is the total number of $i$-species molecules and $\mathcal{V}^{\mathrm{CR}}$ the CR volume, we evaluate $n^{\mathrm{CR}}_{i}$ as:
\begin{equation}
\label{ni}
n^{\mathrm{CR}}_{i}=\frac{1}{\mathcal{V}^{\mathrm{CR}}}\sum_{j=1}^{N_{i}}\theta(z_{j}),
\end{equation}
where:
\begin{equation}
\label{theta}
\theta(z_{j})=
\begin{cases}
1\qquad\text{if}\ z_{j}\in\mathrm{CR}\\
0\qquad\text{otherwise}
\end{cases},
\end{equation}
is a function that selects the molecules inside the CR. In the practice, we let $\theta(z)$ switch continuously to 0 at the CR boundaries, in order to avoid sudden jumps in $n^{\mathrm{CR}}_{i}$.

We then define the function $G(z,Z_{F})$, localizing the action of $F^{\mu}$ close to the force center $Z_{F}$, as:
\begin{equation}
\label{dsigmoid}
G_{w}(z-Z_{F}) = \frac{1}{4w}\left[1+\cosh\left(\frac{z-Z_{F}}{w}\right)\right]^{-1}.
\end{equation}
$G_{w}$ is nonzero only for $z\sim Z_{F}$, with an intensity peak proportional to $w^{-1}$ and a width proportional to $w$. 
We observe that $F^{\mu}$ is non-conservative, since it is not possible to define a potential function that leads to Eq.~(\ref{Fb}).

As a further remark we note that Eq.~(\ref{Fb}) defines two separate forces acting on the solute and solvent species. In principle the concentration of both species affects the chemical potential of the solution \cite{KirkwoodJCP1951} and should be controlled. However, we are mainly interested in constant pressure simulations, where only a single population needs to be subject to $F^{\mu}$, because the action of the barostat algorithm \cite{ParrinelloPRL1980,ParrinelloJAP1981} guarantees a prompt equilibration of the other species concentration. If instead an NVT dynamics is considered, $F^{\mu}$ should act on both solute and solvent species. 

To conclude this section let us summarize the C$\mu$MD scheme in explicit algorithm steps:
\begin{enumerate}
\item the solid-liquid interface position $z_{\mathrm{I}}$ is located on-the-fly analyzing the solvent molecules distribution within the box,
\item the force center $Z_{F}$, is updated maintaining a fixed distance $D_{F}$ from $z_{\mathrm{I}}$, and the CR position is shifted accordingly,
\item the densities $n^{\mathrm{CR}}_{i}$ are evaluated via Eq.~(\ref{ni}).
\item The MD equations of motions, including the external forces of Eq.~(\ref{Fb}), are integrated.
\end{enumerate}
We recall once again that this method is effective for a finite validity time $\tau_{F}$, depending on the availability of molecules in the reservoir.

\section{Test Case and Simulation Setup}\label{simsetup}
\begin{figure}[] 
\centering\includegraphics[scale=0.25]{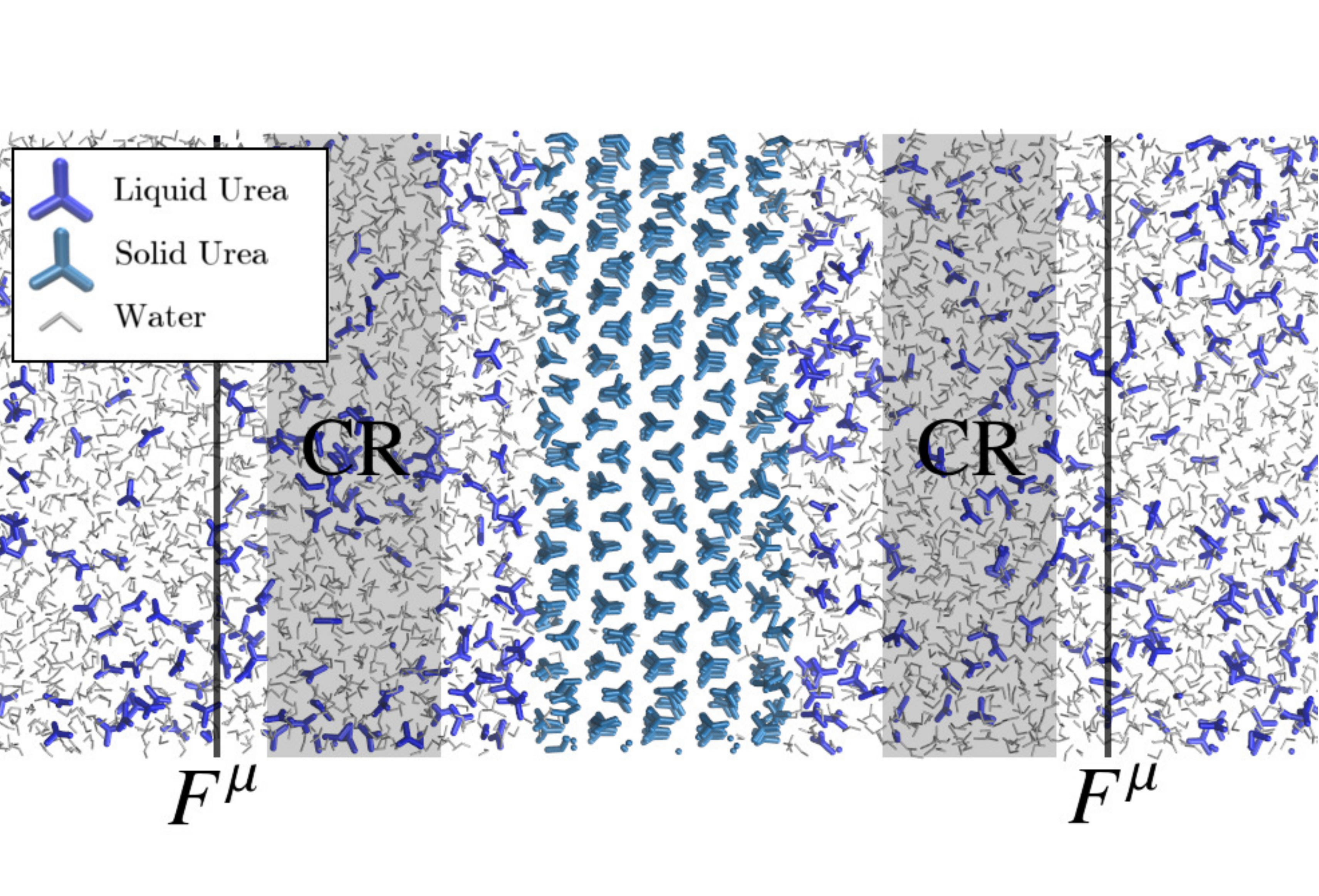}\\  
\caption{\label{box} Typical simulation box for the study of urea crystal growth in aqueous solution. The vertical lines indicate the force center $Z_{F}$, where $F^{\mu}$ is applied (see Eq.~(\ref{Fb})), located at both sides of the slab, at distance $D_{F}$ from the crystal interfaces. The gray-shaded areas indicate the CR.}\end{figure}

In the present section we introduce the typical setup of our calculations.
We considered here the case of a urea crystal growing in aqueous solution in slab geometry, as shown in Fig.~\ref{box}. 
The system is generated starting from an approximately $2.5\,\mathrm{nm}$ thick crystal slab, periodically repeated in the $x$ and $y$ directions. Along $z$, the slab exposes either the $\{001\}$ or the $\{110\}$ face, the two stable faces in urea crystal growth from aqueous solution \cite{DochertyFD1993}.
Such a crystal is immersed in a supersaturated solution, generated by means of the Packmol \cite{MartinezJCC2009} and genbox (GROMACS package) utilities. The typical size of the simulation box is of $2.7\times2.7\times14\,\mathrm{nm}^{3}$, Periodic Boundary Conditions (PBC) are applied in the $x$,$y$ and $z$ directions. The number of urea and of water molecules for each simulated system is reported in Tab.~\ref{tab}. The initial solute concentration has to be larger than the target concentration, to allow a stationary growth regime for a timescale of $10-100$ ns.

Following our previous work \cite{SalvalaglioJACS2012,SalvalaglioAC2013}, we use the Generalized Amber Force Field (GAFF) \cite{CornellJACS1995,WangJCC2004} to model urea, and the TIP3P model \cite{JorgensenJCP1983} for water. The molecular bonds are constrained through the LINCS algorithm \cite{HessJCTC2008b} and long range electrostatics is handled by means of Particle Mesh Ewald (PME) method \cite{DardenJCP1993}. 

The system is kept at a constant temperature of $300$ K and pressure of $1$ bar by using the stochastic velocity rescaling thermostat \cite{BussiJCP2007} and the Parrinello-Rahman barostat \cite{ParrinelloPRL1980}. Because of the planar symmetry, the barostat is applied in its semi-isotropic version, so that the $z$-side of the box, parallel to the growth direction, is rescaled separately from the $x$ and $y$ sides. Since the system is rescaled during the dynamics, also the C$\mu$MD length parameters, as e.g.~$D_{F}$, have to be rescaled accordingly. Our implementation redefines the lengths on-the-fly, following the fluctuations of the $z$-edge of the box. In the considered simulations $F^{\mu}$ does not affect the pressure of the system by a significant amount, since it determines a surface effect, small compared to the potential and kinetic bulk contributions. For this reason in the presented results we did not couple the external force with the barostat equations.

\begin{table*}[]
\centering
\begin{tabular}{c|ccccccccc}
\hline
Run & Face &$N_{\mathrm{u}}$ & $N_{\mathrm{w}}$ & $D_{F}$ [nm] & $\xi$ [nm] & $\sigma$ [nm]&  u/w &$n_{0i}\,[\mathrm{nm}^{-3}]$& $k_{i}\,\left[\frac{\mathrm{nm}^{3}\,\mathrm{kJ}}{\mathrm{mol}}\right]$\\
\hline
$A_{\mathrm{u1}}$ &\multirow{3}{*}{$\{001\}$} & \multirow{3}{*}{$679$} & \multirow{3}{*}{$1757$} & \multirow{3}{*}{$2.5$} & \multirow{3}{*}{$1.0$} & \multirow{3}{*}{$0.5$} & \multirow{3}{*}{u} & $1.5$ & \multirow{3}{*}{$21.0$}\\
$A_{\mathrm{u2}}$ & &&&&&& & $2.4$ & \\
$A_{\mathrm{u3}}$ & &&&&&& & $3.3$ & \\
\hline
$A_{\mathrm{w1}}$ &\multirow{3}{*}{$\{001\}$} & \multirow{3}{*}{$679$} & \multirow{3}{*}{$1757$} & \multirow{3}{*}{$2.5$} & \multirow{3}{*}{$1.0$} & \multirow{3}{*}{$0.5$} & \multirow{3}{*}{w} & $30.25$ & $2.5$ \\
$A_{\mathrm{w2}}$ & & & & & & & & $28.25$ & $2.5$\\
$A_{\mathrm{w3}}$ & & & & & & & & $26.25$ & $3.5$ \\
\hline
NPT & $\{001\}$ & $679$ & $1757$ & \multicolumn{6}{c}{Ordinary NPT} \\
\hline
$B_{\mathrm{u}j}$ & \multirow{2}{*}{$\{110\}$} & \multirow{2}{*}{$625$} & \multirow{2}{*}{$1533$} & \multirow{2}{*}{$3.2$} & \multirow{2}{*}{$1.0$} & \multirow{2}{*}{$0.2$} & u & $1.7$ & $7.8$ \\
$B_{\mathrm{w}j}$ & &&&&&& w & $29.0$ & $1.95$ \\
\hline
NPT & $\{110\}$ &  $625$ & $1533$ & \multicolumn{6}{c}{Ordinary NPT} \\
\hline
\end{tabular}
\caption{\label{tab} Simulation settings used for the different MD calculations presented in Sec.~\ref{growth}.}
\end{table*}
After the initial equilibration phase, a $2$ fs integration step has been chosen for all the production runs. All the calculations have been performed using the GROMACS package \cite{HessJCTC2008} equipped with a private version of the PLUMED plug-in \cite{BonomiCPC2009,TribelloCPC2014}, in which the external force $F^{\mu}$ has been implemented. 

The position $z_{\mathrm{I}}$ of the two interfaces is located on-the-fly by an algorithm which collects the instantaneous water density distribution $n_{\mathrm{w}}(z)$ in an histogram of bin-size $\delta z$. Since  $n_{\mathrm{w}}(z)=0$ inside the crystal volume, we can choose a threshold value $n_{\mathrm{I}}$ that indicates the transition from the liquid phase to the solid. The crystal interfaces are then located at $n_{\mathrm{w}}(z)=n_{\mathrm{I}}$. Typical values used for these parameters are $n_{\mathrm{I}}=10 \mathrm{nm}^{-3}$ and $\delta z=0.75\,\mathrm{\AA}$.

As shown in Fig.~\ref{box}, to comply with the PBC, $F^{\mu}$ is applied at both sides of the slab, and the CR consists of two layers of liquid at fixed distance from the crystal interfaces. All the C$\mu$MD parameters chosen for the presented simulations are based on a preliminary tuning performed on a typical system. During this process the effective decoupling between the reservoir region and the growing crystal is assessed by testing different parameter configurations, and observing the resulting solution behavior. All the details of this tuning procedure are reported in the Supplementary Material (SM). The relevant calculation settings are listed in Tab.~\ref{tab}

\begin{figure*}[] 
\centering\includegraphics[scale=0.4]{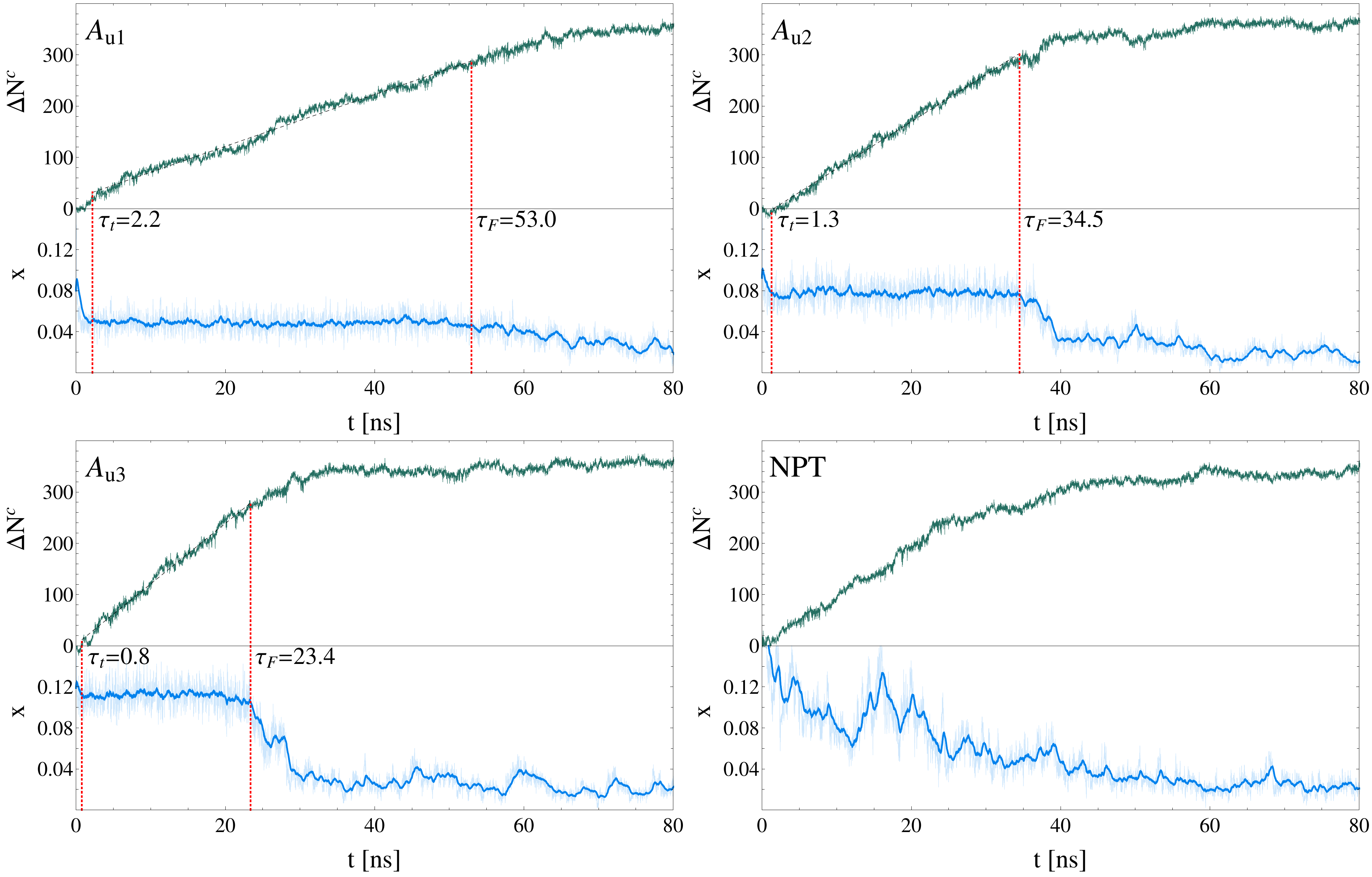}\\  
\caption{\label{g001} 
$A_{\mathrm{u}j}$ simulations results compared to the ordinary NPT behavior. The green curves represent the increase of the crystal-like molecule number $\Delta N^{c}$ as a function of time. The blue curves represent the solution mole fraction $x$ measured in the CR as a function of time. The instantaneous value of $x$ is represented in faded color, while the full-color curves are obtained via Exponentially Weighted Moving Average, with characteristic smoothing time of $0.5$ ns.
The red marks indicate the validity time-window of the C$\mu$MD method, namely $\tau_{\mathrm{t}}<t<\tau_{\mathrm{F}}$. The black dashed lines represent the linear fit of the $\Delta N^{c}$ behavior, calculated within the corresponding validity time range as explained in the text.}\end{figure*}

\section{Crystal Growth at Constant Supersaturation}\label{growth}

We now present the results of the application of the C$\mu$MD method to the simulation of urea crystal growth in aqueous solution. 
In our study we have considered the growth of the $\{001\}$ and $\{110\}$ faces. It is known that these two faces are characterized by different growth mechanisms: the $\{001\}$ face undergoes a rough growth process, while the $\{110\}$ face grows through a birth-and-spread mechanism \cite{PianaNat2005,SalvalaglioJACS2012}.

First, we investigate the $\{001\}$ face growth process. 
We have performed 2 sets of three simulations, referred to as $A_{\mathrm{u}j}$ and $A_{\mathrm{w}j}$. In the $A_{\mathrm{u}j}$ type simulations $F^{\mu}$ restrains urea density, while in the $A_{\mathrm{w}j}$ is water to be controlled. As reported in Tab.~\ref{tab}, the index $j=1,2,3$ runs over different target densities.
For comparison we have also performed an ordinary NPT run of the same system. In each simulation we have evaluated the number of solid molecules $N^{c}$ using the Degree Of Crystallinity (DOC) variable defined in Ref.~\onlinecite{GibertiCES2014}.

\begin{figure}[] 
\centering\includegraphics[scale=0.4]{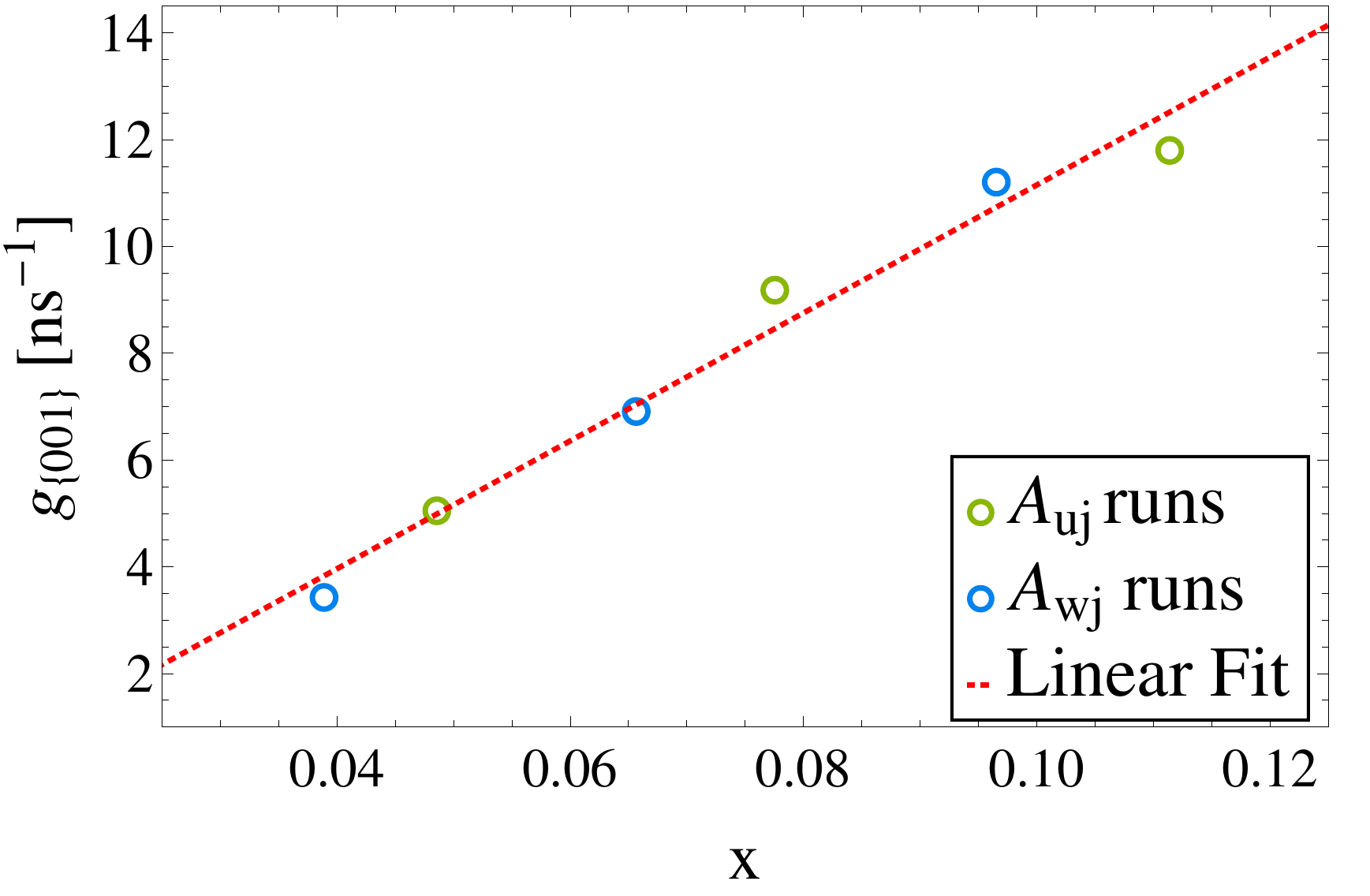}\\  
\caption{\label{grate} Growth rates measured in $A_{\mathrm{u}j}$ and $A_{\mathrm{w}j}$ simulations, versus the average CR mole fraction (see Fig.~\ref{g001}). The red dashed line is a linear fit of the growth rates.}\end{figure}

\begin{figure}[] 
\centering\includegraphics[scale=0.43]{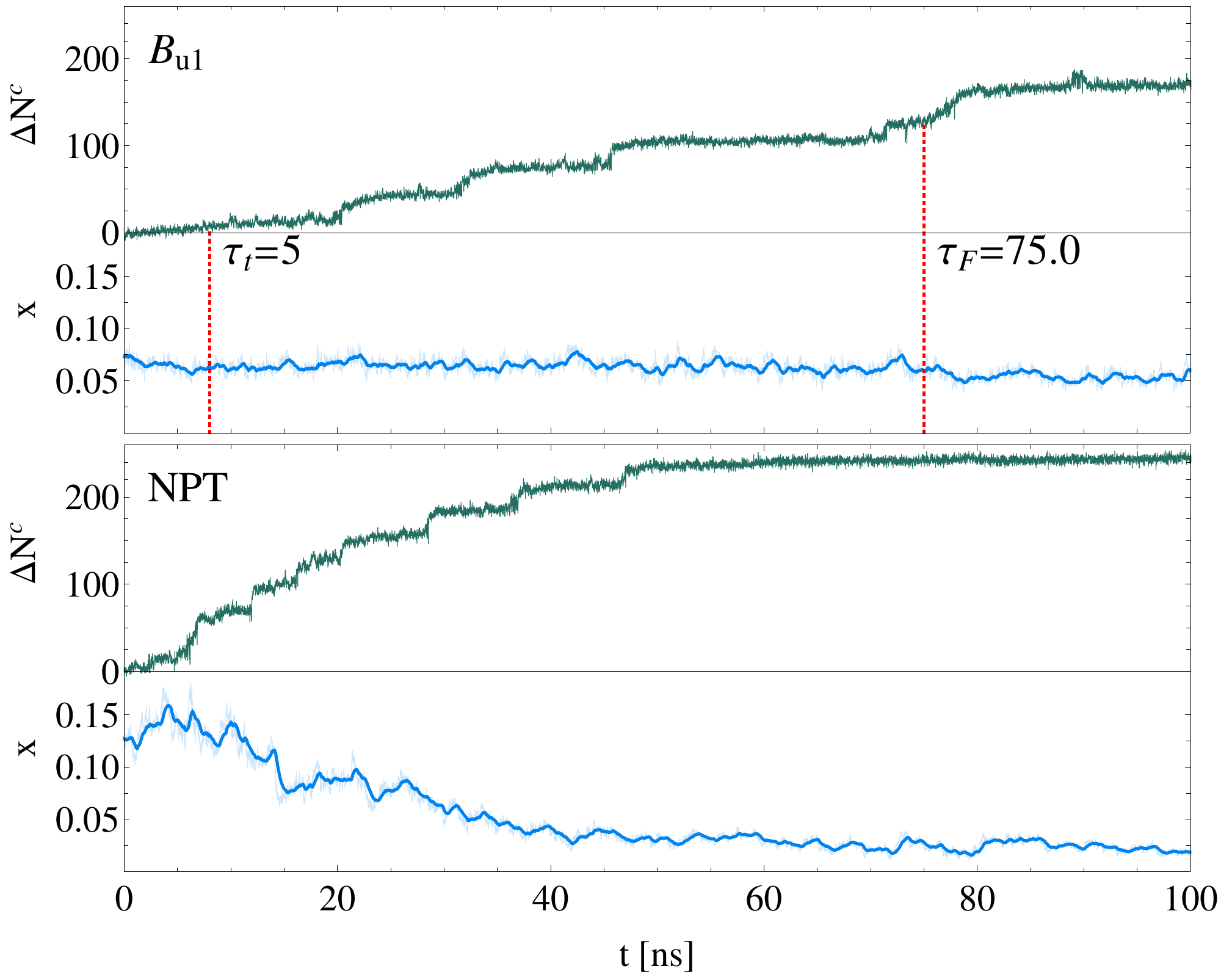}\\  
\caption{\label{g110} $B_{\mathrm{u}1}$ simulations results compared to the ordinary NPT behavior. The green curves represent the increase of the crystal-like molecule number $\Delta N^{c}$ as a function of time. The blue curves represent the solution mole fraction $x$ measured in the CR as a function of time. The instantaneous value of $x$ is represented in faded color, while the full-color curves are obtained via Exponentially Weighted Moving Average, with characteristic smoothing time of $0.5$ ns.
The red marks indicate the validity time-window of the C$\mu$MD method.}
\end{figure}
In Fig.~\ref{g001} we report the evolution of $N^{c}$ and of the mole fraction $x$ measured in the CR for the $A_{\mathrm{u}j}$ and NPT simulations (an analogous plot for the $A_{\mathrm{w}j}$ runs is included in the SM). Let us focus on the behavior of the $A_{\mathrm{u1}}$ simulation (upper-left panel). After an initial transient $\tau_{\mathrm{t}}$ the action of $F^{\mu}$ stabilizes the mole fraction around $x=0.05$. As the crystal grows, the reservoir region is gradually depleted and, as a consequence, at $\tau_{F}=53$ ns the CR mole fraction starts drifting to lower values, meaning that the C$\mu$MD method is no longer valid. 
Within the validity range ($\tau_{\mathrm{t}}<t<\tau_{\mathrm{F}}$) $N^{c}$ increases linearly, as expected for a rough growth process at constant supersaturation. The $N^{c}$ behavior during the validity range can be fitted with a linear model, the slope of which is the growth rate $g_{\{001\}}$ at this particular solution composition. For $t>\tau_{F}$ the crystal rate decreases, reflecting the change in solution chemical potential. An analogous behavior is obtained in the $A_{\mathrm{u}2}$ and $A_{\mathrm{u}3}$ simulations, where larger supersaturations are enforced, determining a faster crystal growth. This results in a more rapid depletion of the reservoir and, as a consequence, in a shorter $\tau_{F}$. In contrast with the C$\mu$MD runs, in the NPT simulation the solution concentration varies throughout the whole dynamics, interfering with crystal growth. 
In Fig.~\ref{grate} we report the growth rates obtained in the $A_{\mathrm{u}j}$ and $A_{\mathrm{w}j}$ simulations, corresponding to different values of $x$. Remarkably, $g_{\{001\}}$ exhibits a linear dependence on $x$ that is characteristic of a rough growth mechanism (see e.g.~Ref.~\onlinecite{LovetteIECR2008}). It is rewarding that the results are consistent, irrespective from the controlled species.

In two further sets of simulations we have applied C$\mu$MD method to the $\{110\}$ face growth, either restraining urea ($B_{\mathrm{u}}$ run in Tab.~\ref{tab}) or water density ($B_{\mathrm{w}}$). For comparison we have also performed an ordinary NPT simulation of the same system. 
As mentioned before, the $\{110\}$ face growth process exhibits a birth-and-spread nature. Because of this mechanism $N^{c}$ evolves in a step-wise behavior, each step corresponding to the formation of a new crystalline layer. The dynamics of $N^{c}$ for the $B_{\mathrm{u}}$ and NPT cases is represented in Fig.~\ref{g110}. In the figure we have also reported the mole fraction evolution, showing that our method is able to maintain a stable solution composition for a significant time. In both $B_{\mathrm{u}}$ and $B_{\mathrm{w}}$ runs the CR mole fraction is maintained at approximately $x=0.062$ (all the $B_{\mathrm{u}}$ and $B_{\mathrm{w}}$ results are reported in the SM) .

We now estimate the $\{110\}$ face growth rate at constant supersaturation $g_{\{110\}}$. Since the environment solution is not changing, we can assume that after a layer growth event the system carries no memory of its past. Thus the probability that a new layer is created is independent from the growth history, and the time interval between two successive events follows an exponential distribution\cite{ChkoniaJCP2009}. 
However, in a single run, only few layers are created before the solution starts depleting. Thus, in order to collect sufficient statistics for the construction of the time distribution, we have repeated each $B_{\mathrm{u}}$ and $B_{\mathrm{w}}$ simulations five times. Of course we have considered only the events occurring at $\tau_{\mathrm{t}}<t<\tau_{\mathrm{F}}$.

\begin{figure}[] 
\centering\includegraphics[scale=0.3]{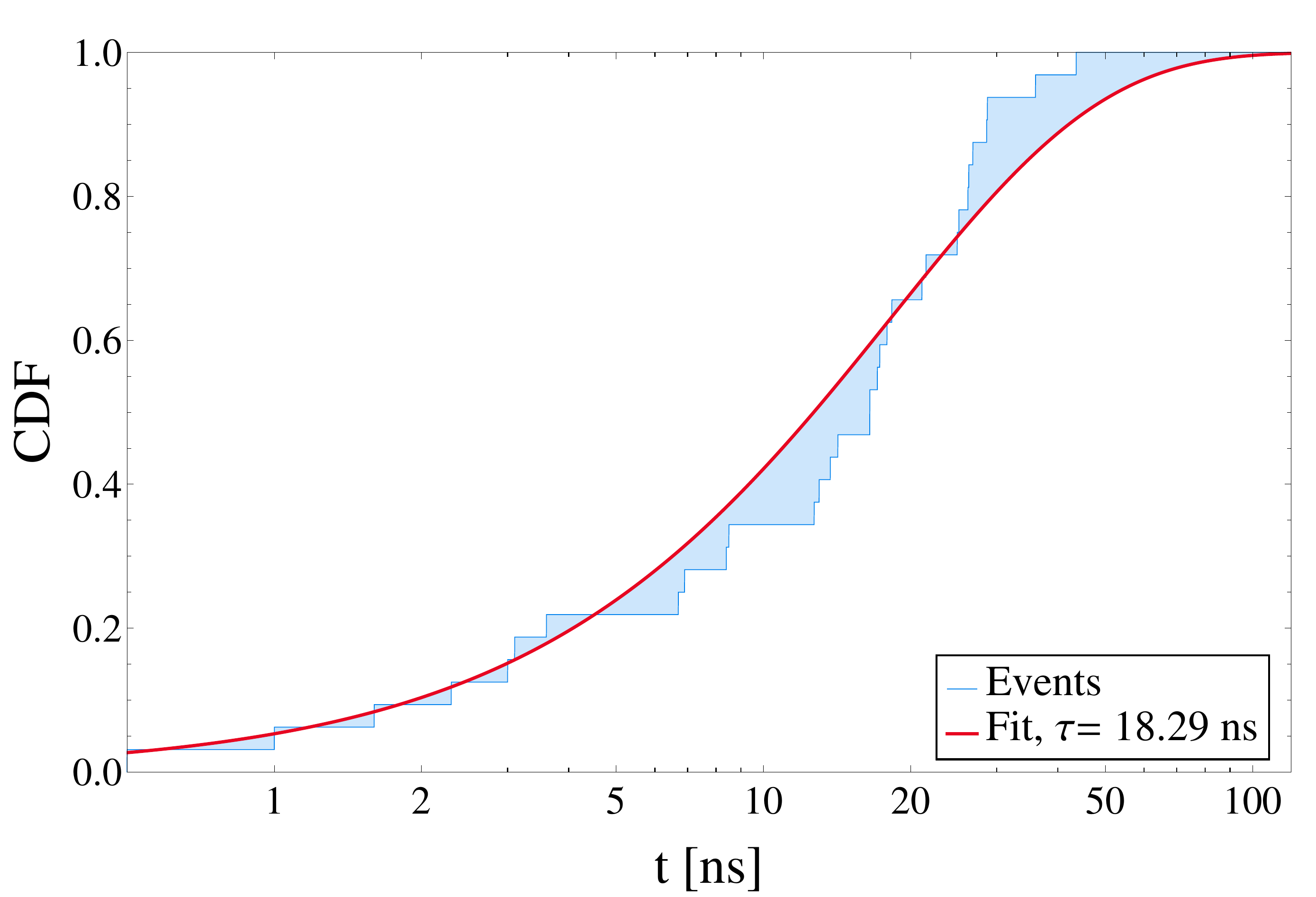}\\  
\caption{\label{r110} Cumulative time distribution of the layer growth events at $x=0.062$ ($B_{\mathrm{u}}$ and $B_{\mathrm{w}}$ simulations). The red curve is the fitted Cumulative Distribution Function, equal to $y(t)=1-\exp(-t/\tau_{\{110\}})$ for exponentially distributed time intervals.}\end{figure}

We have constructed the cumulative time distribution of the observed events, and fitted it with an exponential model, as shown in Fig.~\ref{r110}. The statistical significance of this fit has been assessed using the Kolmogorov-Smirnov test \cite{MasseyJASA1951}, obtaining a p-value of 0.37. From the regression we have extracted the characteristic occurrence time $\tau_{\{110\}}=18.3\pm2.7$ ns, which corresponds to a growth rate $g_{\{110\}}=1.73\pm0.25\,\mathrm{ns}^{-1}$. This result can be compared to our estimate of the $\{001\}$ rate at $x=0.062$, that is $g_{\{001\}}=6.59\pm0.97\,\mathrm{ns}^{-1}$, obtained from the linear interpolation in Fig.~\ref{grate}. 

In Ref.~\onlinecite{SalvalaglioAC2013} we have proposed a model for the growth rate of the $\{hjk\}$ face of urea crystal, in which:
\begin{equation}
\label{angrate}
g_{\{hjk\}}=g_{0}\exp\left(-\beta\Delta G_{\{hjk\}}\right),
\end{equation}
where $g_{0}$ is a characteristic diffusion limited growth rate, $\Delta G_{\{hjk\}}$ is the free-energy barrier associated to the creation of an extra $\{hjk\}$ layer and $\beta^{-1}=k_{\mathrm{B}}T$. The model has been used in Ref.~\onlinecite{SalvalaglioAC2013} to predict the urea crystal habits, by evaluating the velocity ratio between different crystal faces:
\begin{equation}
\label{ratio}
g_{\{hjk\}}/g_{\{lmn\}}=\exp(-\beta\Delta G_{\{hjk\}}+\beta\Delta G_{\{lmn\}}).
\end{equation}
We underline that this ratio depends only on the free-energy barriers, which are static properties of the system. Using the $\Delta G_{\{001\}}$ and $\Delta G_{\{110\}}$ evaluated via Well-Tempered (WT) metadynamics\cite{BarducciPRL2008} in Ref.~\onlinecite{SalvalaglioAC2013}, Eq.~(\ref{ratio}) gives $g_{\{001\}}/g_{\{110\}}=3.61$ at $x=0.062$. This result is consistent with the ratio obtained from our dynamical estimates, that is $g_{\{001\}}/g_{\{110\}}=3.8\pm0.8$. 

The birth-and-spread nature of the $\{110\}$ face growth mechanism is determined by non-negligible free-energy barriers $\Delta G_{i}$ associated to the incorporation of new solute molecules in the growing crystal. In the following we shall derive such free-energy barriers from the C$\mu$MD trajectories obtained from the $B_{\mathrm{u}}$ and $B_{\mathrm{w}}$ simulations.

\begin{figure}[] 
\centering\includegraphics[scale=0.3]{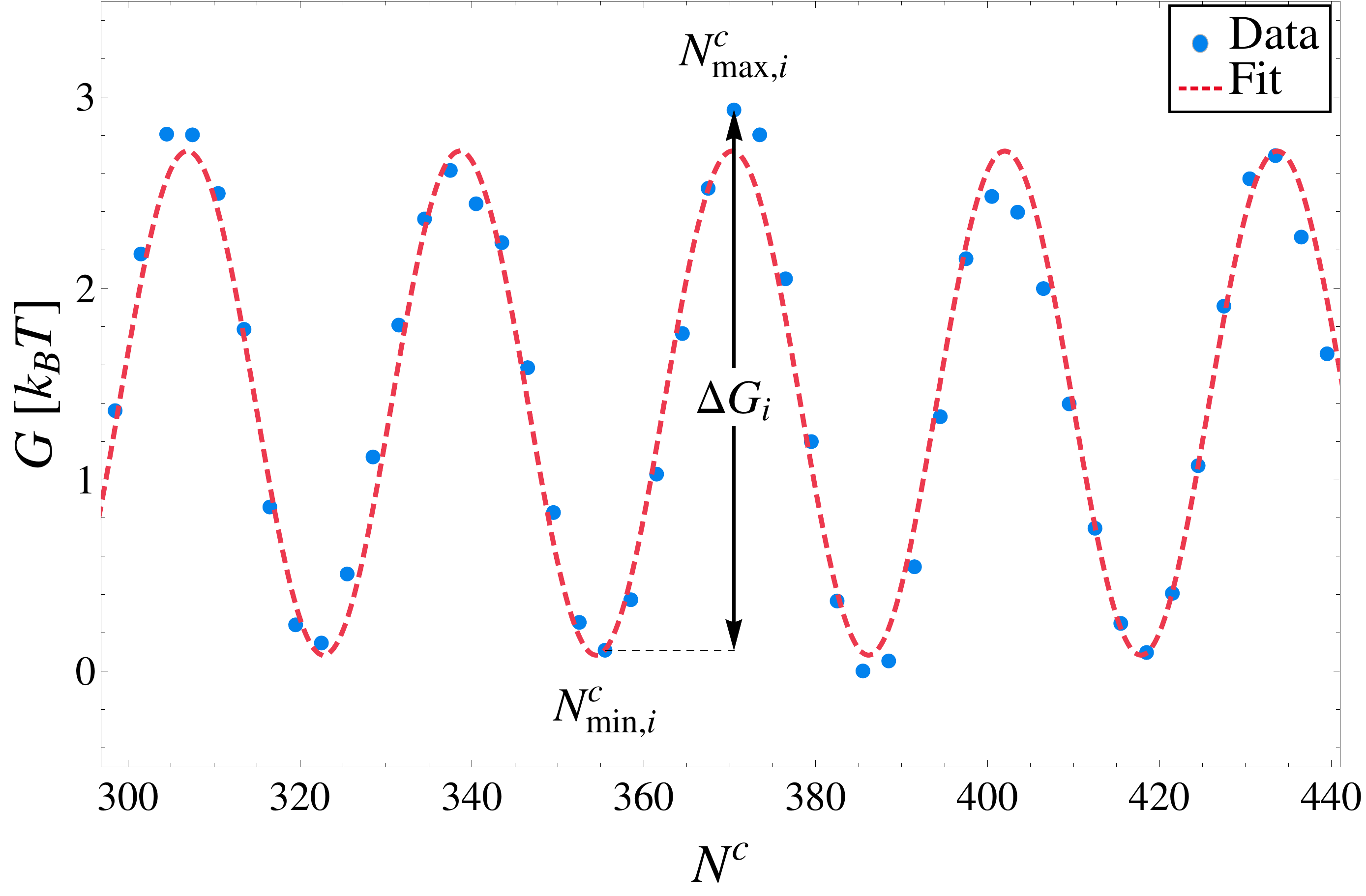}\\ 
\caption{\label{FESg} Behavior of $G(N^{c})$ (see Eq.~(\ref{FES})) evaluated using the probability distribution $P(N^{c})$, extracted from the $B_{\mathrm{u}}$ and $B_{\mathrm{w}}$ simulations sampling. The red dashed line represents a sinusoidal fit of the numerical result.}
\end{figure}

The growth dynamics, see e.g.~Fig.~\ref{g110}, shows that the system evolves through a sequence of metastable states $i=1\ldots n$, each characterised by an average number of molecules in the crystal state $\langle{N^{c}}\rangle_i$. Therefore, the probability distribution $P(N^{c})$, computed averaging over the ensemble of $B_{\mathrm{u}}$ and $B_{\mathrm{w}}$ trajectories, will exhibit an alternation of minima and maxima, the former representing the metastable states at $N_{\mathrm{min},i}^{c}=\langle{N^{c}}\rangle_i$, and the latter representing the faster layer growth transitions. This is in agreement with Fig.~\ref{FESg}, which represents the behavior of:
\begin{equation}
\label{FES}
G(N^{c})\equiv-\frac{1}{\beta}\ln P(N^{c}).
\end{equation}
From the function $G(N^{c})$ we can evaluate the free-energy barriers associated to the layer growth transitions. Since the lifetime of each metastable state is such that the sampling in the vicinity of a minimum can be considered ergodic, the relative free energy associated to the $N^{c}$ fluctuations within the $i$-th basin can be written as:
\begin{equation}
\label{DG1}
\Delta{G}_{N^{c}_{\mathrm{min},i}\rightarrow{N^{c}}}=-\frac{1}{\beta}\ln \frac{P(N^{c})}{P(N^{c}_{\mathrm{min},i})}=G(N^{c})-G(N^{c}_{\mathrm{min},i})
\end{equation}
According to Eq.~(\ref{DG1}), the free-energy barrier associated to the creation of a layer from a metastable state is $\Delta G_{i}=G(N_{\mathrm{max},i}^{c})-G(N^{c}_{\mathrm{min},i})$, as indicated in Fig.~\ref{FESg}.

The barriers resulting from our C$\mu$MD simulations exhibit very similar heights, showing that successive growth events obtained in the molecular model are equivalent. This provides further confirmation that the action of C$\mu$MD maintains the growing crystal environment at a constant chemical potential. 
As shown in Fig.~\ref{FESg} the data can be fitted using a sinusoid, obtaining a remarkable agreement. The free-energy barrier associated to the sampled growth events can be extracted from the fitted curve, obtaining $\Delta G=2.63\, k_{\mathrm{B}}T$, which is in substantial agreement with the corresponding barrier $\Delta G=2.0\pm1.0\, k_{\mathrm{B}}T$, computed in the model proposed in Ref.~\onlinecite{SalvalaglioAC2013}.

\section{Conclusions and Perspectives}\label{concl}
In this work we have presented the C$\mu$MD method, which addresses the concentration depletion problems in the MD study of finite systems. 
C$\mu$MD applies an external force that controls the solution composition within a region of interest of the system, while the remaining volume is used as a molecular reservoir. 

C$\mu$MD has been implemented and tested for the relevant case of urea crystal growth in aqueous solution. For the considered systems the method is capable of maintaining the solution environment of the growing crystal at constant composition, up to times of the order of $10\div100$ ns. This has allowed us to estimate the $\{001\}$ and $\{110\}$ faces growth rates in a constant supersaturation environment. Such kind of results are  not accessible with ordinary NPT dynamics, due to the intrinsic coupling between the crystal size and the solution composition. The retrieved results are also in remarkable agreement with the predictions of Ref.~\onlinecite{SalvalaglioAC2013}, which are based on the combination of a theoretical free energy model and WT metadynamics calculations.
The evaluation of the free-energy barriers associated to the growth of consecutive $\{110\}$ layers has shown that the events are thermodynamically equivalent, providing further support to the hypothesis that C$\mu$MD can establish a stationary growth regime. 

The proposed scheme can be applied to a variety of MD problems, such as crystal nucleation, electro-chemistry, surfactants adsorption. However the external force has to be properly re-defined to comply with the characteristic features of these systems. 
The timescale of validity of C$\mu$MD is related to the fixed number of molecules involved in the simulation. A possible future development can be the combination of C$\mu$MD method with adaptive resolution \cite{WangPRX2013,PotestioPRL2013} or particle insertion techniques \cite{HongJCP2012}, which would result in a considerable extension of the validity range of the method.

\pagebreak

\onecolumngrid

\newpage
\begin{center}
\textbf{\large Molecular Dynamics Simulations of Solutions at Constant Chemical Potential\\ Supplementary Material}
\end{center}
\setcounter{section}{0}
\setcounter{equation}{0}
\setcounter{figure}{0}
\setcounter{table}{0}
\setcounter{page}{1}
\makeatletter
\renewcommand{\thesection}{S\Roman{section}}
\renewcommand{\theequation}{S\arabic{equation}}
\renewcommand{\thefigure}{S\arabic{figure}}
\renewcommand{\thetable}{S\Roman{table}}
\renewcommand{\bibnumfmt}[1]{[S#1]}
\renewcommand{\citenumfont}[1]{S#1}
\vspace*{1cm}
In the following we report further details about the results presented in the Main Submission (MS) text. This supplementary material is organized in two sections: Sec.~\ref{ub1} presents a preliminary study that has been performed to tune the C$\mu$MD method for the typical system considered in our calculations. In Sec.~\ref{growthSI} we gather supplementary information concerning the simulation results presented in the MS.
\vspace*{1cm}
\twocolumngrid

\section{Preliminary Study}\label{ub1}
In this section we report the results of a preliminary study on the application of C$\mu$MD method to crystal growth. This study aims at finding convenient parameter settings for the typical urea-water systems simulated in our work.

It has to be noted that C$\mu$MD reliability is related to system properties such as size, initial solution concentration and liquid diffusivity. For this reason, a tuning process becomes necessary each time the method is applied to a new system.

Let us first summarize the C$\mu$MD parameters:
\begin{itemize}
\item $D_{F}$, the distance of the force center $Z_{F}$ from the solid-liquid interface $z_{\mathrm{I}}$.
\item the CR position and size, defined by $\xi$ and $\sigma$. The former is the distance between the inner CR boundary and $z_{\mathrm{I}}$, while the latter is the distance between the outer CR boundary and $Z_{F}$ (see Fig.~2 in the MS). 
\item the force constant $k_{i}$ (see MS Eq.~(1)),
\item the characteristic length $w$ of MS Eq.~(4),
\item the target concentration $n_{0i}$, either for urea or water species.
\end{itemize}

\begin{table*}[]
\centering
\begin{tabular}{c|cccccccc}
\hline
Run & $N_{\mathrm{u}}$ & $N_{\mathrm{w}}$ & $D_{F}$ [nm] & $\xi$ [nm] & $\sigma$ [nm]&  $w$ [nm] &$n_{\mathrm{0u}}\,[\mathrm{nm}^{-3}]$& $k_{\mathrm{u}}\,\left[\frac{\mathrm{nm}^{3}\,\mathrm{kJ}}{\mathrm{mol}}\right]$\\
\hline
$X_{1}$ & \multirow{4}{*}{$679$} & \multirow{4}{*}{$1757$} & $2.0$ & \multirow{4}{*}{$1.0$} & $0.0$ & \multirow{4}{*}{$0.2$} & \multirow{4}{*}{$1.5$} & \multirow{4}{*}{$21.0$}\\
$X_{2}$ & & & $2.5$ & & $0.5$ & & &\\
$X_{3}$ & & & $3.0$ & & $1.0$ & & &\\
$X_{4}$ & & & $3.5$ & & $1.5$ & & &\\
\hline
NPT$_{2V}$ & $698$ & $5125$ & \multicolumn{6}{c}{Ordinary NPT, double size} \\
\hline
$Y_{1}$ & \multirow{4}{*}{$679$} & \multirow{4}{*}{$1757$} & \multirow{4}{*}{$2.5$} & \multirow{4}{*}{$1.0$} & \multirow{4}{*}{$0.5$} & \multirow{4}{*}{$0.2$} & \multirow{4}{*}{$1.5$} & $35.0$ \\
$Y_{2}$ & & & & & & & & $21.0$\\
$Y_{3}$ & & & & & & & & $7.0$\\
$Y_{4}$ & & & & & & & & $3.5$\\

\hline
\end{tabular}
\caption{\label{tabSI} Simulation settings used for the MD simulations reported in this section.}
\end{table*}
The system used for this study is the $\{001\}$ growth configuration, all the relevant parameters are summarized in Tab.~\ref{tabSI}. In particular, the following analysis is focused on tuning of $D_{F}$, $\sigma$ and $k_{\mathrm{u}}$, while constant values have been assigned to $n_{\mathrm{u}0}$, $\xi$ and $w$.  In order to keep a limited growth rate, and thus a long C$\mu$MD validity time, $n_{\mathrm{u}0}$ has been set to $1.5\,\mathrm{nm}^{-3}$. The choice for the $\xi$ and $w$ values will be discussed later on.

\begin{figure*}[] 
\centering\includegraphics[scale=0.4]{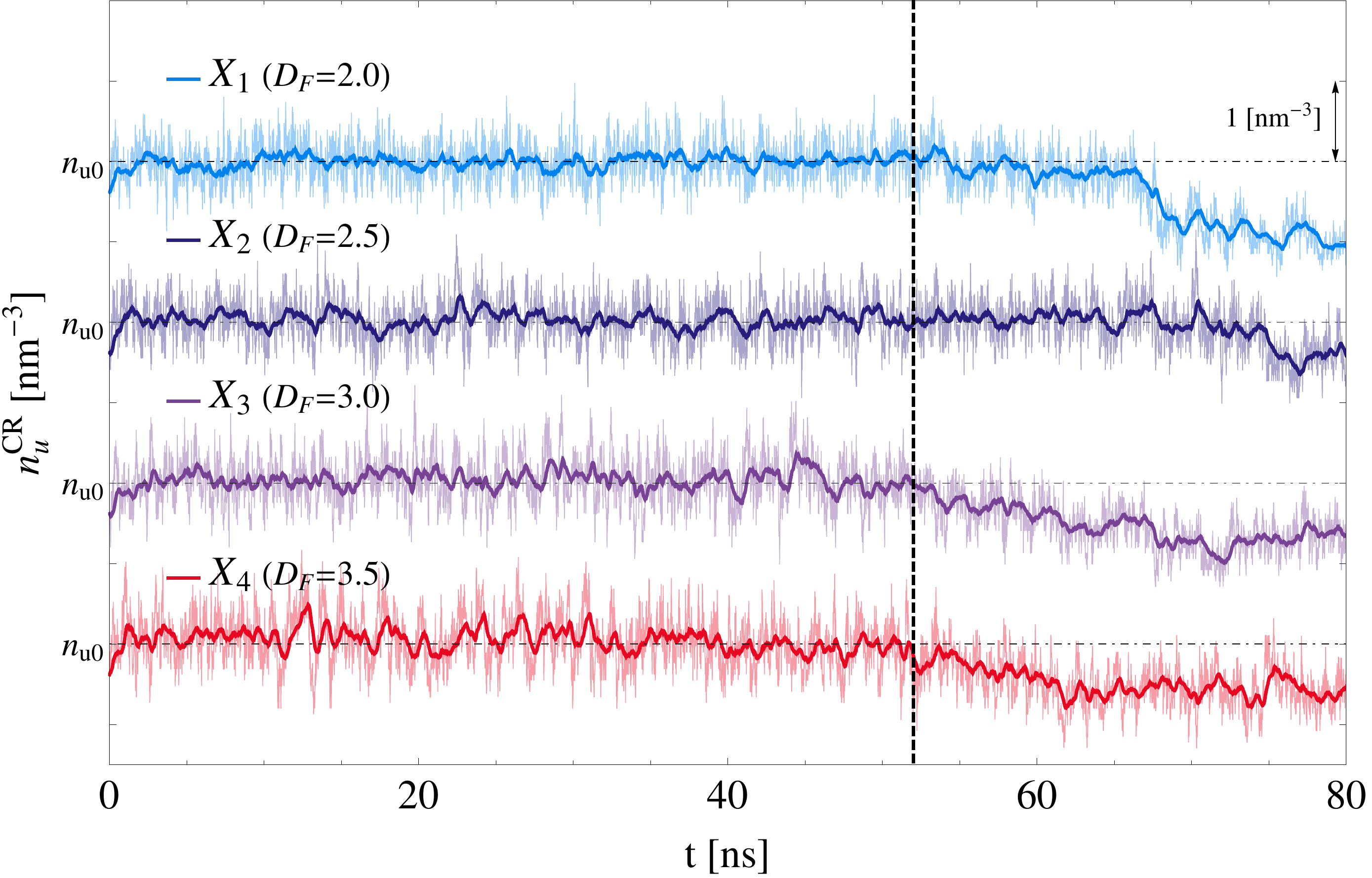}\\  
\caption{\label{fluctd} Trajectories of the CR urea density for the $X_{j}$ simulations. The instantaneous value is represented in faded color, while the full-color curves are obtained via Exponentially Weighted Moving Average, with characteristic smoothing time of $0.5$ ns. Each curve is referred to the target density $n_{0\mathrm{u}}=1.5\,\mathrm{nm}^{-3}$ (dot-dashed axes). The dashed vertical line indicates $t=52$ ns, the time at which solution depletion becomes significant for $X_{3}$ and $X_{4}$.}\end{figure*}

In a first set of 4 MD runs, which we will refer to as $X_{j}$, we have tested the effect of varying $D_{F}$ and $\sigma$ together, so that the force is applied at different distances from the crystal interface while the CR maintains the same size and position (referred to the crystal). 

In Fig.~\ref{fluctd} we report the evolution of the CR urea density $n^{\mathrm{CR}}_{\mathrm{u}}$. It can be noted that, in each simulation, the C$\mu$MD maintains a stable $n^{\mathrm{CR}}_{\mathrm{u}}$, close to the target value $n_{\mathrm{u0}}$, for at least $52$ ns ($X_{3}$ and $X_{4}$). In the $X_{1}$ and $X_{2}$ cases, for which $F^{\mu}$ is applied closer to the CR, the validity time is longer. This is because a larger volume is assigned to the reservoir region, and the control of the solution concentration is more effective. 

To look at the quantitative differences between Fig.~\ref{fluctd} curves, in Tab.~\ref{tab2} we report the average CR density and standard deviation for each run, evaluated over $52$ ns. During this lapse of time all the simulations enforce an average density close to the target value $n_{\mathrm{u}0}$, and a correlation between the density fluctuations and $\sigma$ (or, alternatively, $D_{F}$) emerges. That is because a larger distance between the force center and the CR results in a slower response of $F^{\mu}$ to the density variations, and thus in a weaker control of $n^{\mathrm{CR}}_{\mathrm{u}}$. 
\begin{table}[]
\centering
\begin{tabular}{c|cc}
\hline
Run & $\langle n^{\mathrm{CR}}_{\mathrm{u}}\rangle-n_{\mathrm{u0}}$ & $s_{n}$ \\
\hline
$X_{1}$ & $-2.0\times10^{-3}$ &$0.24$\\
$X_{2}$ & $3.1\times10^{-2}$ &$0.26$\\
$X_{3}$ & $3.8\times10^{-2}$ &$0.30$\\
$X_{4}$ & $4.9\times10^{-2}$ &$0.33$\\
\hline
\end{tabular}
\caption{\label{tab2} Average CR urea density (referred to $n_{\mathrm{u0}}=1.5$) and standard deviation $s_{n}=\sqrt{\langle (n^{\mathrm{CR}}_{\mathrm{u}})^{2}\rangle-\langle n^{\mathrm{CR}}_{\mathrm{u}}\rangle^{2}}$ of the $X_{j}$ simulations. The density, in units of $\mathrm{nm}^{-3}$, is averaged over $52$ ns of dynamics, in order to consider only the validity regime of the method.}
\end{table}
\begin{figure}[] 
\centering\includegraphics[scale=0.3]{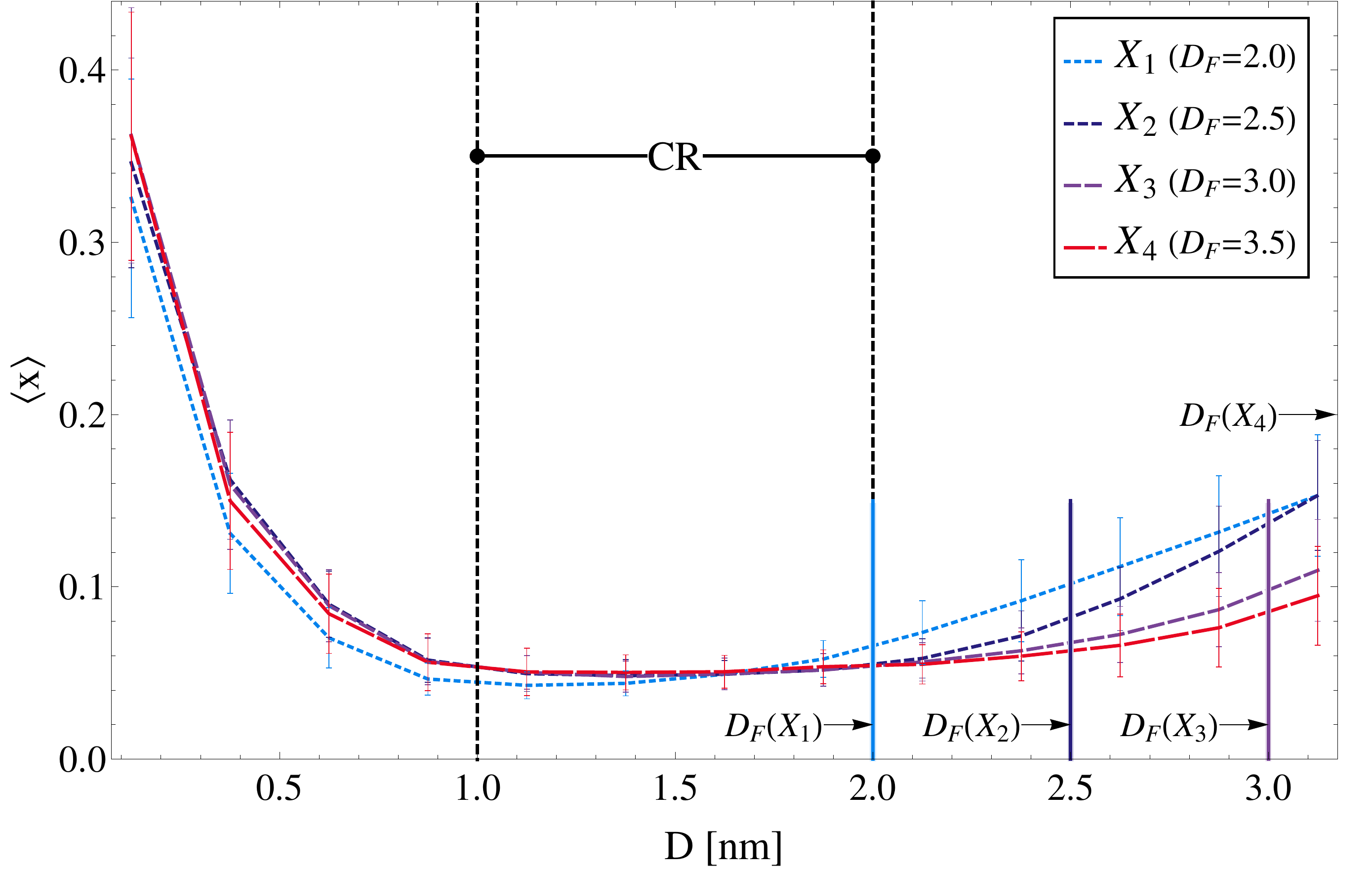}\\  
\caption{\label{xprof} Local mole fraction profile as a function of the distance from the solid-liquid interface. The results of the $X_{j}$ simulations, averaged over $\tau=41$ ns, are displayed. Each local value is evaluated in $0.25$ nm thick layers, symmetrically located at both sides of the crystal. The vertical colored lines highlight the force center position except for the $X_{4}$ case, in which $D_{F}$ falls outside the plot range. The error bars are evaluated with the re-blocking technique \cite{FlyvbjergJCP1989}. Because of the long correlation times the error estimate is not completely reliable for $D\gtrsim 2$ nm.}
\end{figure}
\begin{figure}[] 
\centering\includegraphics[scale=0.42]{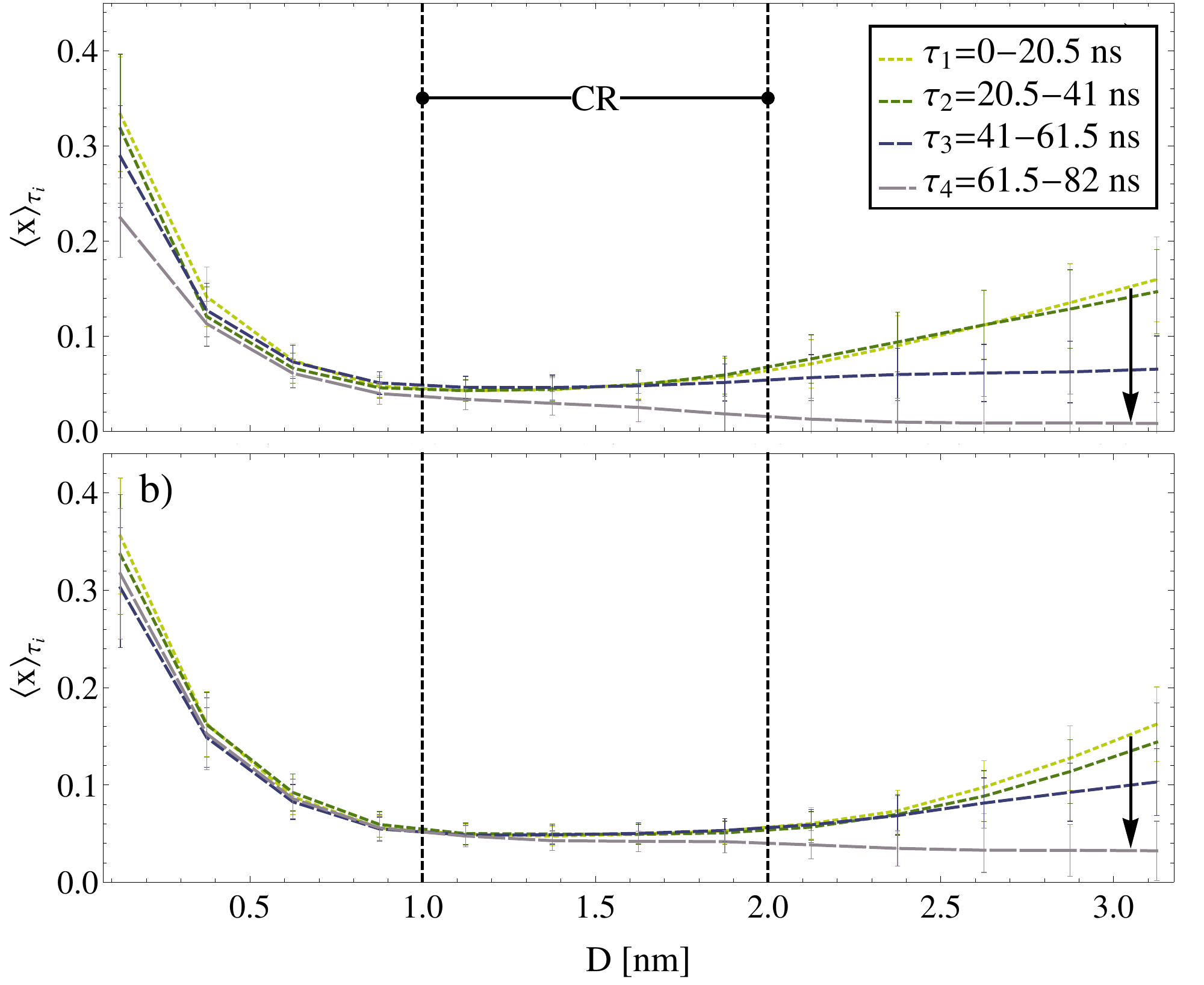}\\  
\caption{\label{xproft} Temporal evolution of the average mole fraction profile for the $X_{1}$ (a) and $X_{2}$ (b) cases. Each curve represents the average of $x(D)$ over a different time window, as indicated in the legend. The black arrows highlight the temporal evolution of the composition in the reservoir. The error bars are evaluated with the re-blocking technique \cite{FlyvbjergJCP1989}. Because of the long correlation times the error estimate is not completely reliable for $D\gtrsim 2$ nm.}
\end{figure}

In the $X_{j}$ runs we have also monitored the mole fraction at different distances from the crystal. In Fig.~\ref{xprof} we report the resulting mole fraction profile for each simulation, averaged in time. The qualitative behavior of this profile is consistent with the trend described in Sec.~2 of the MS and, in particular, we underline that the choice of $\xi=1.0$ nm is appropriate to identify the TR of our system.

Fig.~\ref{xprof} shows that, for the $X_{2}$, $X_{3}$ and $X_{4}$ cases, the application of C$\mu$MD method determines a flat profile within the CR, and no correlation with $D_{F}$ is present. 
Conversely, in the $X_{1}$ case the vicinity of the force center to the CR has a direct contribution on its concentration. This should be avoided, since it prevents the decoupling of the reservoir region from the environment of the crystal.
To support this statement we compare the time evolution of the $x$ profile for the $X_{1}$ and $X_{2}$ cases. To this purpose we have evaluated the time average of the $x$ profiles over 4 consecutive windows of time, so that the slow composition dynamics is visible. Let us focus on the behavior of $x$ within the CR region: while in the $X_{1}$ case (represented in Fig.~\ref{xproft}a) the profile varies throughout the whole simulation, in the $X_{2}$ case (Fig.~\ref{xproft}b) the mole fraction remains flat for the first three windows of time, up to the limit of validity of C$\mu$MD.

These results suggest that a layer of thickness $\sigma$ between the force center and the CR is necessary to decouple the reservoir from the crystal environment. However, the choice of $\sigma$ also affects the promptness of $F^{\mu}$ in balancing the CR composition variations. Given this, the configuration used in $X_{2}$ case seems to be the best compromise. We underline that the choice of $\sigma$ is also related to the characteristic length $w$, which indicates the localization of $F^{\mu}$ around the force center. In our calculations we have chosen $w=0.2\div0.3$ which, according to our experience, determines an efficient, but still local, action on the molecules.

In a further set of 4 simulations, indicated in Tab.~\ref{tabSI} as $Y_{j}$, we have performed C$\mu$MD with different values of the force constant $k_{\mathrm{u}}$, starting from the $X_{2}$ settings. As a term of comparison we have also performed an ordinary NPT simulation, named NPT$_{2V}$, in which the crystal slab is immersed in a larger volume of liquid, almost twice the original size, with a solution composition close to that enforced by $F^{\mu}$ in the $Y_{j}$ runs. 

\begin{figure*}[] 
\centering\includegraphics[scale=0.4]{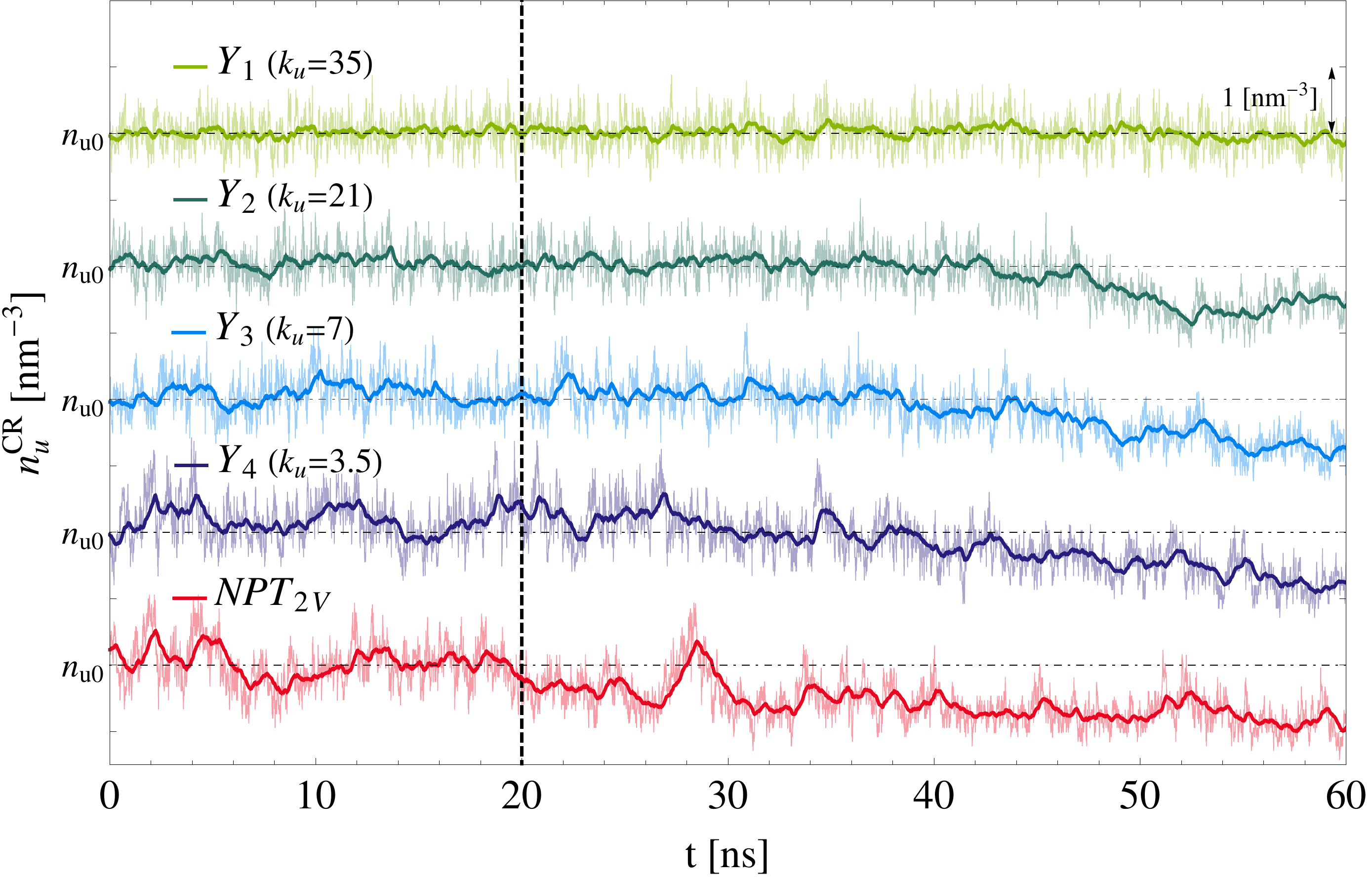}\\  
\caption{\label{fluctk} Trajectories of the CR urea density for the $Y_{j}$ and NPT$_{2V}$ simulations. The instantaneous value is represented in faded color, while the full-color curves are obtained via Exponentially Weighted Moving Average, with characteristic smoothing time of $0.5$ ns. Each curve is referred to $n_{0\mathrm{u}}=1.5\,\mathrm{nm}^{-3}$ (dot-dashed axes). The dashed vertical line indicates $t=20$ ns, when solution depletion becomes significant in the NPT$_{2V}$ run.}\end{figure*}

In Fig.~\ref{fluctk} we report the evolution of the CR urea concentration $n^{\mathrm{CR}}_{\mathrm{u}}$ in the $Y_{j}$ and NPT$_{2V}$ simulations. First we note that the NPT$_{2V}$ system does not experience a significant solution depletion up to $20$ ns. Thus, within this time range, we can consider the NPT$_{2V}$ run as a good approximation to a macroscopic system. 

If we look to the $Y_{j}$ trajectories we see that the $n^{\mathrm{CR}}_{\mathrm{u}}$ dynamics is affected by the choice of $k$. In the $Y_{1}$ case ($k_{\mathrm{u}}=35.0\,\mathrm{nm}^{3}\,\mathrm{kJ}/\mathrm{mol}$) $n^{\mathrm{CR}}_{\mathrm{u}}$ fluctuates steadily around $n_{\mathrm{u0}}$, while for smaller $k_{\mathrm{u}}$ the density is less stable.
This is quantitatively confirmed by Tab.~\ref{tab3}, where we report the corresponding $n^{\mathrm{CR}}_{\mathrm{u}}$ mean and standard deviations, up to $20$ ns. As $k$ is reduced, the standard deviation increases, while the average density distances the target value $n_{\mathrm{u0}}$.
That is because a larger $k_{\mathrm{u}}$ determines a more intense $F^{\mu}$ response to the density fluctuations, and thus a more effective control on the flux of urea molecules from the reservoir to the CR.

From Tab.~\ref{tab3} we also note that, while a larger $k$ allows a better control of the CR composition, the density fluctuations in the realistic NPT dynamics are more similar to the results of $Y_{4}$ simulation, the one with the smallest $k$. This raises the question whether this difference in composition sampling affects the crystal growth. 

\begin{table}[]
\centering
\begin{tabular}{c|cc}
\hline
Run & $\langle n^{\mathrm{CR}}_{\mathrm{u}}\rangle-n_{\mathrm{u0}}$ & $s_{n}$ \\
\hline
$Y_{1}$ & $1.5\times10^{-2}$ &$0.24$\\
$Y_{2}$ & $5.5\times10^{-2}$ &$0.26$\\
$Y_{3}$ & $7.5\times10^{-2}$ &$0.29$\\
$Y_{4}$ & $1.8\times10^{-1}$ &$0.34$\\
NPT$_{2V}$ & $1.6\times10^{-2}$ &$0.34$\\
\hline
\end{tabular}
\caption{\label{tab3} Average CR urea density (referred to $n_{\mathrm{u0}}=1.5$) and standard deviation of the $Y_{j}$ and NPT$_{2V}$ simulations. The density, in units of $\mathrm{nm}^{-3}$, is averaged over $20$ ns of dynamics.}
\end{table}

\begin{figure}[] 
\centering\includegraphics[scale=0.3]{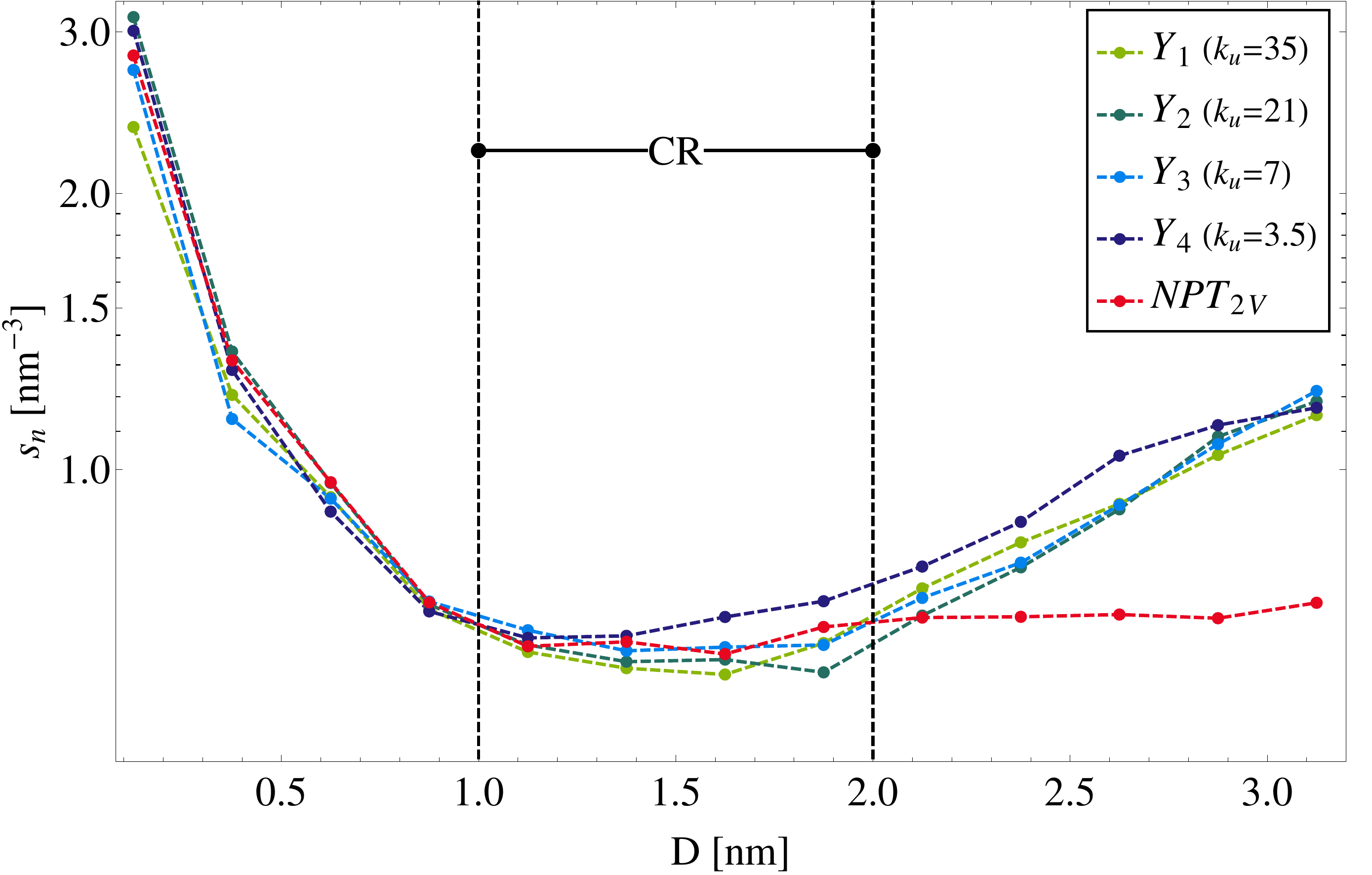}\\  
\caption{\label{sigmak} Logarithmic plot of the $n_{\mathrm{u}}^{\mathrm{CR}}$ standard deviation, for the $Y_{j}$ and NPT$_{2V}$ runs. Each value is measured by detecting $n_{\mathrm{u}}$ within a $0.25$ nm thick layer placed at a distance $D$ from the solid-liquid interface. The dashed vertical lines indicate the CR boundaries.}\end{figure}

To answer this question in Fig.~\ref{sigmak} we report the local standard deviation $s_{n}(D)$ of $n^{\mathrm{CR}}_{\mathrm{u}}$. While at distances $D\geq 1.5$ nm from the interface the $Y_{j}$ runs exhibit a different $s_{n}$ behavior compared to the NPT$_{2V}$ case, this difference is much mitigated as we get closer to the solid-liquid interface. This analysis suggests that, for this system, the choice of $k_{\mathrm{u}}$ does not affect significantly the composition fluctuations of the solution in contact with the crystal. As a result it is preferable to choose a large $k_{\mathrm{u}}$, which entails a more effective control on the CR.

As a last remark we underline that in this study $F^{\mu}$ was applied exclusively to the urea specie. However the results are still useful when the restraining force acts on the water population, provided that $k$ is rescaled according to the typical $n_{\mathrm{w}}$ fluctuations.
\section{Crystal Growth Calculations}\label{growthSI}
In this section we report the additional material concerning the MD calculations presented in Sec.~4 of the MS. We organize the information in two sub-sections, concerning the growth processes of the $\{001\}$ and $\{110\}$ faces of urea.

\subsection{$\mathbf{\{001\}}$ face growth.}
\begin{figure*}[] 
\centering\includegraphics[scale=0.4]{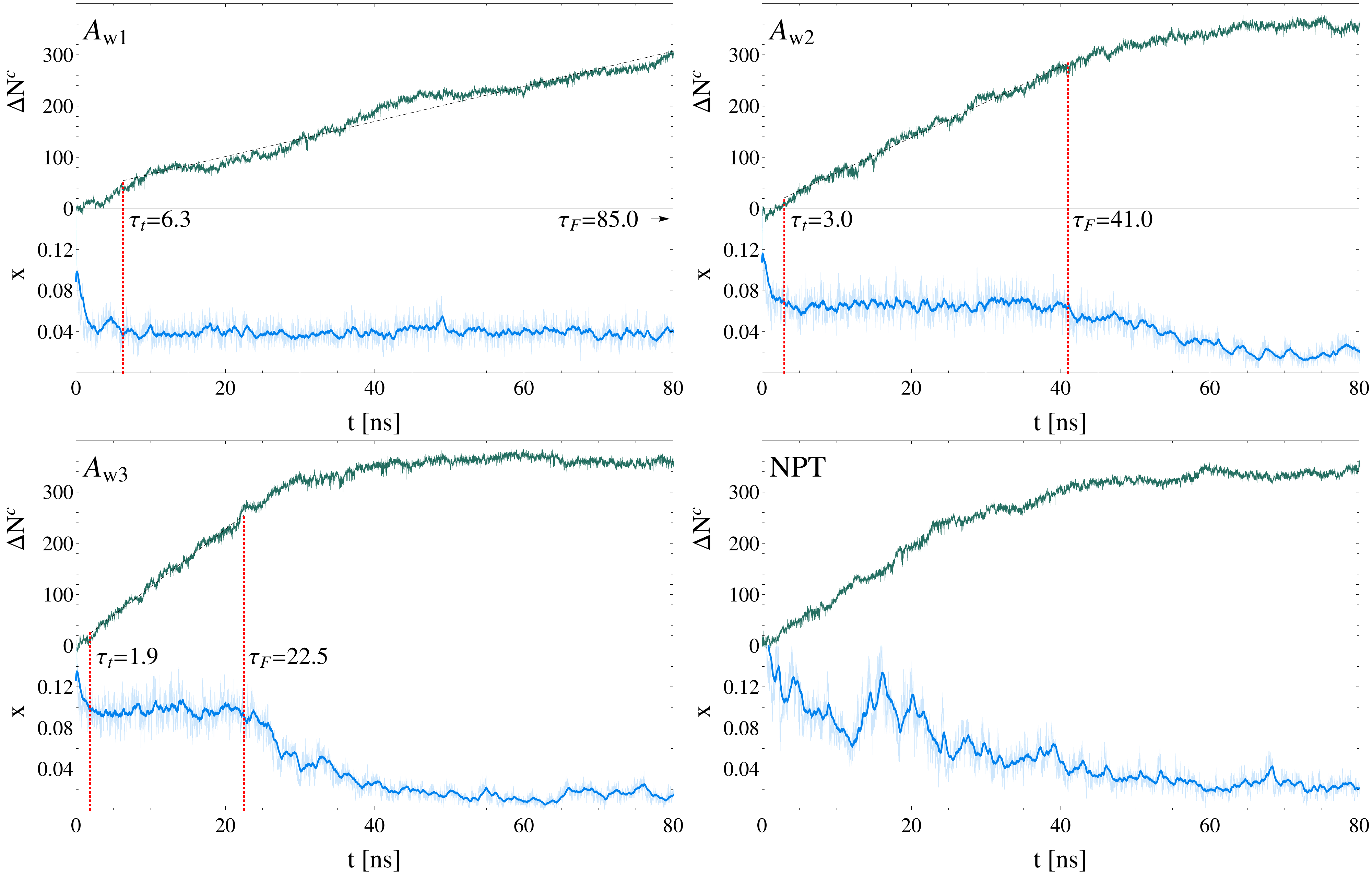}\\  
\caption{\label{g001SI} 
$A_{\mathrm{w}j}$ simulations results compared to the ordinary NPT behavior. The green curves represent the increase of the crystal-like molecule number $\Delta N^{c}$ as a function of time. The blue curves represent the solution mole fraction $x$ measured in the CR as a function of time. The instantaneous value of $x$ is represented in faded color, while the full-color curves are obtained via Exponentially Weighted Moving Average, with characteristic smoothing time of $0.5$ ns.
The red marks indicate the validity time-window of the C$\mu$MD method, namely $\tau_{\mathrm{t}}<t<\tau_{\mathrm{F}}$. The black dashed lines represent the linear fit of the $\Delta N^{c}$ behavior, calculated within the corresponding validity time range.}\end{figure*}

For the study of $\{001\}$ face growth we have performed two sets of simulations, named $A_{\mathrm{u}j}$ and $A_{\mathrm{w}j}$ (see Tab.~I of MS). In the first one the urea population is controlled, while in the second water is controlled. The results of the $A_{\mathrm{u}j}$ set are reported in the MS, here we show the results of the $A_{\mathrm{w}j}$ runs,  in Fig.~\ref{g001SI}, compared to the ordinary NPT dynamics.
Also in this set of simulations the C$\mu$MD method succeeds in establishing a linear growth regime, even though the control of $x$ is less rigid and allows larger fluctuations.

The dependence of the growth rates extracted from both the $A_{\mathrm{u}j}$ and $A_{\mathrm{w}j}$ simulations on the corresponding solution mole fraction has been assessed with a linear fit, represented in Fig.~5 of the MS. The parameters resulting from the interpolation are the following:
\begin{eqnarray}
 q &=& -0.83\pm0.75\,\mathrm{ns}^{-1},  \nonumber \\
 m &=& 119.8\pm9.6\,\mathrm{ns}^{-1}, \nonumber
\end{eqnarray}
where the linear model is $y=mx+q$.

\subsection{$\mathbf{\{110\}}$ face growth.}
\begin{figure*}[]
\centering\includegraphics[scale=0.45]{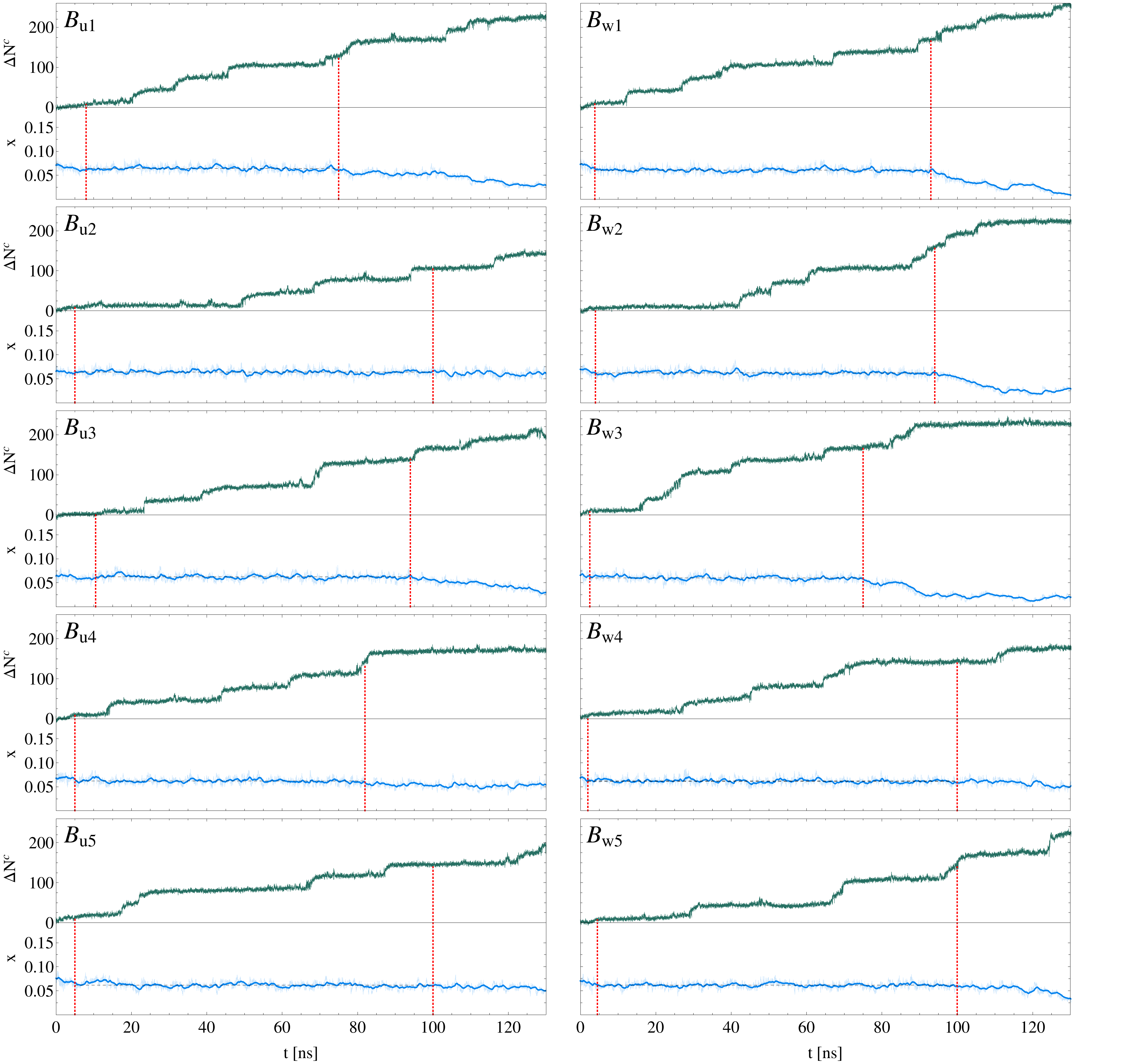}\\  
\caption{\label{g110SI}  $B_{\mathrm{u}j}$ and $B_{\mathrm{w}j}$ simulations results. The green curves represent the increase of the crystal-like molecule number $\Delta N^{c}$ as a function of time. The blue curves represent the solution mole fraction $x$ measured in the CR as a function of time. The instantaneous value of $x$ is represented in faded color, while the full-color curves are obtained via Exponentially Weighted Moving Average, with characteristic smoothing time of $0.5$ ns. The horizontal dashed line shows the average of $x$, evaluated over the validity time-window of the C$\mu$MD method, indicated by red marks.}\end{figure*}

For the study of $\{110\}$ face growth the $B_{\mathrm{u}j}$ and $B_{\mathrm{w}j}$ simulation sets have been performed, the relative parameters are shown in Tab.~I of MS. Each set is constituted by 5 repetitions of the same simulations, in order to collect a relevant statistics for the presented analysis. 
Moreover, the C$\mu$MD settings have been designed so that both $B_{\mathrm{u}j}$ and $B_{\mathrm{w}j}$ simulations reach analogous growth conditions. For this reason, in our analysis, we have considered the results of the $B_{\mathrm{u}j}$ and $B_{\mathrm{w}j}$ runs as equivalent. In the multiple plot of Fig.~\ref{g110SI} we report the resulting dynamics of each simulation.

In the MS we have reported the estimate of the $\{110\}$ face growth rate, obtained from the trajectories displayed in Fig.~\ref{g110SI}. As explained in the text, under constant supersaturation conditions we can assume that the probability of growth events occurrence follows a Poisson distribution. Therefore the waiting time between two consecutive events is exponentially distributed, and the growth rate can be extracted from this distribution.
To this purpose the time intervals between two consecutive growth events have been collected. We have considered only those events occurring within the validity time of C$\mu$MD, during which the growth could be considered at constant supersaturation. We have also discarded the waiting time preceding the first event, since it is affected by the initial transient time, during which the solution is brought at the desired composition.

From the resulting statistics we have then constructed the cumulative time distribution, and fitted it with the cumulative distribution associated to an exponential probability density:
\begin{equation}
\label{expmod}
y(t)=1-\exp(-t/\tau_{\{110\}}).
\end{equation}
From the fit we have extracted $\tau_{\{110\}}=18.3\pm2.7$ ns (as shown in Fig.~7 of the MS). The corresponding error has been obtained with the bootstrap technique \cite{EfronAoS1979}. The growth rate has been estimated by calculating the average $N^{c}$ increase associated to the growth events ($31.6\pm0.3$), and dividing it by $\tau_{\{110\}}$, to obtain the result reported in the MS.


%
%

%

\end{document}